\documentclass[aps,prb,amsmath, amssymb, english, twocolumn,reprint,floatfix, superscriptaddress]{revtex4-1}

\usepackage{multirow}
\usepackage{float}
\usepackage[normalem]{ulem}
\usepackage{textgreek}

\usepackage{bbm}
\usepackage{comment}
\usepackage{soul}

\usepackage{tikz}
\usepackage{microtype}
\usepackage{tkz-euclide}
\usetkzobj{all}
\usepackage{tkz-graph}
\usepackage{epstopdf}
\usepackage{epsfig} 
\usepackage[applemac]{inputenx} 
\usepackage{amsmath}

\usepackage[unicode]{hyperref}
\hypersetup{
 colorlinks,
 citecolor=blue,
 filecolor=blue,
 linkcolor=blue,
 urlcolor=blue
}

\newcommand{\ket}[1]{\left|#1\right>}
\newcommand{\bra}[1]{\left<#1\right|}

\newcommand{\cc}{c^{\phantom{\dagger}}}

\newcommand{\cd}{c^\dag}

\newcommand{\dd}{d^\dag}

\graphicspath{{./Images/}}

\date{}
\begin{document}
\title{Driven-dissipative quantum mechanics on a lattice: Simulating a fermionic reservoir on a quantum computer}
\author{Lorenzo Del Re} 
\affiliation{Department of Physics, Georgetown University, 37th and O Sts., NW, Washington,
DC 20057, USA}
\affiliation{Erwin Schr\"odinger International Insitute for Mathematics and Physics, Boltzmanngasse 9
1090 Vienna, Austria}
\author{Brian Rost} 
\affiliation{Department of Physics, Georgetown University, 37th and O Sts., NW, Washington,
DC 20057, USA}
\author{A. F. Kemper} 
\affiliation{Department of Physics, North Carolina State University, Raleigh, North Carolina 27695, USA}
\author{J. K. Freericks} 
\affiliation{Department of Physics, Georgetown University, 37th and O Sts., NW, Washington,
DC 20057, USA}
\date{\today} 


\begin{abstract}
The driven-dissipative many-body problem remains one of the most challenging unsolved problems in quantum mechanics. The advent of quantum computers may provide a unique platform for efficiently simulating such driven-dissipative systems. But there are many choices for how one can engineer the reservoir. One can simply employ ancilla qubits to act as a reservoir and then digitally simulate them via algorithmic cooling. A more attractive approach, which allows one to simulate an infinite reservoir, is to integrate out the bath degrees of freedom and describe the driven-dissipative system via a master equation, that can also be simulated on a quantum computer. In this work, we consider the particular case of non-interacting electrons on a lattice driven by an electric field and coupled to a fermionic thermostat.  Then, we provide two different quantum circuits:  the first one reconstructs the full dynamics of the system using Trotter steps, while the second one dissipatively prepares the final  non-equilibrium steady state in a single step. We run both circuits on the IBM quantum experience. For circuit (i), we achieved up to 5 Trotter steps. When partial resets become available on quantum computers, we expect that the maximum simulation time can be significantly increased. The methods developed here suggest generalizations that can be applied to simulating interacting driven-dissipative systems.
\end{abstract}
\maketitle
\section{Introduction}
Dissipation is ubiquitous in nature and often a many-body system of interest is coupled to other degrees of freedom that play the role of an external reservoir (such as electrons and phonons in solid-state physics). The understanding of dissipative many-body quantum systems represents a long standing problem that traces back to the seminal works by Caldeira and Leggett \cite{Caldeira1983a,Caldeira1983b,Leggett1987}, and has experienced a renewed interest in the last decade. In fact, dissipation has been theoretically proposed as a resource for quantum computation \cite{Verstraete2009,Diehl2008,Diehl2010,Yi2012} and experimentally it has been demonstrated that an open quantum system can employ dissipation for quantum state preparation \cite{Barreiro2011,Krauter2011,Lin2013}. Our interest, however, is motivated by the advent of recent pump-probe experiments (see Refs.~\onlinecite{pump-probe-review,pump-probe-review2} for recent reviews), where systems can be easily driven out of equilibrium and then probed at different time delays to determine how they relax. Condensed matter systems always have electrons coupled to a phonon reservoir. Hence, we are ultimately interested in the possibility of eventually using quantum computers to simulate driven-dissipative (and strongly interacting) systems of fermions coupled to bosons.

\par In addition to removing energy from a system, dissipation can also be the source of the phenomena one wants to study. For example, in the case of lattice electrons driven by an electric field, an isolated noninteracting system displays Bloch oscillations\cite{Bloch1929,Zener1934}, leading to an alternating current due to Bragg reflection at the Brillouin zone boundaries. Conversely, when interactions are turned on, dynamical mean field theory (DMFT) \cite{Georges1996} predicts that Bloch oscillations are damped as the system heats to an infinite temperature steady state where the current ultimately vanishes \cite{Freericks2006}.
When dissipation with the environment is taken into account, both noninteracting\cite{Han2012} and interacting \cite{Amaricci2012,Li2015} systems stabilize a DC current in the steady state, that depends non trivially on the interactions and the electric field intensities. Of course, this is exactly what any real system also does, as we know from Ohm's law. Properly treating the dissipation is critical to being able to understand physical phenomena like Ohm's law.

\par The difficulty in addressing strongly correlated systems in a nonperturbative way remains an obstacle for the classical simulation of driven-dissipative systems, even for model systems like the Hubbard model. However, simulations of quantum many-body models have been successfully performed using cold atom quantum simulators \cite{Bloch2008}. Such simulations are analog simulations, meaning that a physical system ({e.g.} cold atoms), set under specific physical conditions ({e.g.} placed in an optical lattice), can reproduce the dynamics of another physical system of interest ({e.g.} electrons in solids). 
Other quantum simulators include trapped ions, which intrinsically simulate the transverse-field Ising model with tunable long-range interactions \cite{Kim2010,Britton2012}, or the Dicke model~\cite{bollinger_dicke}. One can view these quantum simulators as essentially just being well-controlled experiments, which does allow one to learn new things about these complex systems. Nevertheless, there is great interest in transitioning toward digital quantum computation, following the progression from analog to digital classical computation.

\par Some progress has already been made in this realm, although the fact that current hardware is noisy (so-called noisy intermediate-scale quantum computers or NISQ machines), makes it quite difficult to perform time evolution accurately. For example, some simplifications for the Heisenberg model on small clusters allowed the time-evolution to be simulated without needing to Trotterize the evolution operators~\cite{heisenberg1,akhil}. In addition, there exist robust algorithms for time-evolving quantum computers when fault-tolerant quantum computing becomes available~\cite{childs}. An alternative approach to exact time evolution is to evolve systems variationally\cite{variational-evolution}, which is likely to also be a robust approach for NISQ-era machines.

\par There is a fair amount of work that has been completed already on how to simulate open quantum systems on quantum computers. Ref.~\onlinecite{Barreiro2011} showed how to prepare entangled states by simulating a master equation with a digital quantum circuit, whose dissipative nonunitary ``gates'' were obtained by resetting ancilla qubits that had been suitably entangled with the system qubits. For most systems of interest, the size of the bath is much larger than that of the system, often taken to be infinite. In these cases, if one could simulate a master equation,  it would be more convenient than digitally implementing the unitary dynamics of the system plus a finite bath (so-called algorithmic cooling\cite{algorithmic-cooling}). This is because a direct representation of the bath requires far too many qubits, or accuracy is sacrificed to reduce the bath to a reasonable size. However, the master equation approach integrates out the bath's degrees-of-freedom and replaces them with operations that act only on the system. This makes it possible, in principle, to accurately simulate interactions with arbitrarily large baths without requiring arbitrarily large resources.

\par However, when dealing with master equations, approximations are usually necessary. One common choice is to use the Redfield Master Equation (RME) which is obtained by making the Born-Markov approximation. This places some constraints on the applicability of the method. Furthermore, it is not always possible to simulate the RME on a quantum computer. This is because the RME is not guaranteed to generate a quantum dynamical semigroup \cite{Breuer2002}, and further approximations may be needed. 
\par In the first part of our paper, we discuss these approximations thoroughly for the case of an exactly solvable model and compare our results to the exact solution \cite{Han2012}. In particular, we consider a tight-binding model of fermions with periodic boundary conditions (PBC) driven by an electric field and interacting with an external (fermionic) reservoir.

\par In the second part of the paper, we show how to engineer quantum circuits that reproduce the dynamics of this dissipative system. In particular, we devise two schemes: 
\begin{enumerate}
    \item[(i)]  We simulate the dynamics of the Trotterized system directly on a quantum machine computer;
    \item[(ii)] We show how to dissipatively prepare the long-time non-equilibrium steady state (NESS) in a single step.
\end{enumerate}
For (ii) we need to know in advance the state that we want to prepare. For this simple system we can do this analytically, but in general we could also determine it from using (i). This is useful to reduce the number of quantum gates needed to prepare the steady state allowing us to perform quantum operations on the NESS.
We implemented two quantum circuits, corresponding to schemes (i) and (ii), using IBMQ and obtain good results. For scheme (i), we could only run a few Trotter steps because IBMQ (and most current quantum computers) does not have the capability to reset ancilla qubits while leaving the others undisturbed. This forces us to use SWAP operations and extra qubits to accomplish a reset. Better results are expected once partial reset gates become available. 

In this work we provide some important ideas that will hopefully help in the effort to devise quantum algorithms for simulating more complicated and realistic driven-dissipative systems, which we discuss further in the conclusion. We hope this work will provide a stepping stone toward tackling the simulation of more common solid-state systems (e.g. electrons interacting with a phonon bath). Because the problem of studying driven-dissipative systems is simultaneously important and challenging, we feel that our exposition of a concrete approach to a simple system will be useful for both considering more complex scenarios as well as gaining a general understanding of the complex phenomena that driven-dissipative quantum systems may exhibit.

In Sec.~\ref{sec:II}, we first derive the RME of our model and, using an additional approximation, derive a master equation in Lindblad form which is more suitable for simulation on a quantum computer. We compare these approximations with the exact solution and between each other. 
In Sec.~\ref{sec:III}, we explicitly derive the Kraus maps for the two different schemes (i) and (ii). 
In Sec.~\ref{sec:IV}, we  construct the quantum circuits for solving the dynamics of the system (i), and for the state preparation of the NESS (ii). Finally, we show data obtained directly from an IBM quantum computer.
In Sec.~\ref{sec:V}, we summarize the main achievements of our work. 

\section{The  Model}\label{sec:II}
The open system that we consider is given by noninteracting lattice fermions on a one-dimensional chain with nearest-neighbor hopping in the presence of an electric field. The effect of the electric field is taken into account by introducing a complex Peierls phase \cite{Turkowski2005} $\varphi(t) = \Omega\,t$ (given by $\Omega = eEa$) to the hopping integral $\gamma$; we use $\gamma$ for the hopping instead of the more common $t$, so as to not confuse the hopping term with time. The system Hamiltonian then reads:
\begin{equation}\label{ham:1}
\hat{\mathcal H} = -\gamma\sum_{i}e^{i\varphi(t)}d^{\dag}_{i}d^{\phantom{\dagger}}_{i+1} + \mbox{h.~c.}
\end{equation}
Every site of the chain is coupled to an independent infinite fermionic bath, whose Hamiltonian is $\hat{\mathcal H}_b = \sum_{i\alpha}\omega_\alpha\cd_{i\alpha}\cc_{i\alpha}$, through a bilinear hybridization term that is given by:
\begin{equation}\label{ham:2}
\hat{\mathcal V} = -g\sum_{i\alpha} \dd_i \cc_{i\alpha} + \mbox{h.~c.}.
\end{equation} 
Here $g$ is the bare interaction strength, and $\alpha$ is an index that runs over all the internal degrees of freedom of the bath, which are taken to be infinite.
\begin{figure}
\includegraphics[width = \columnwidth]{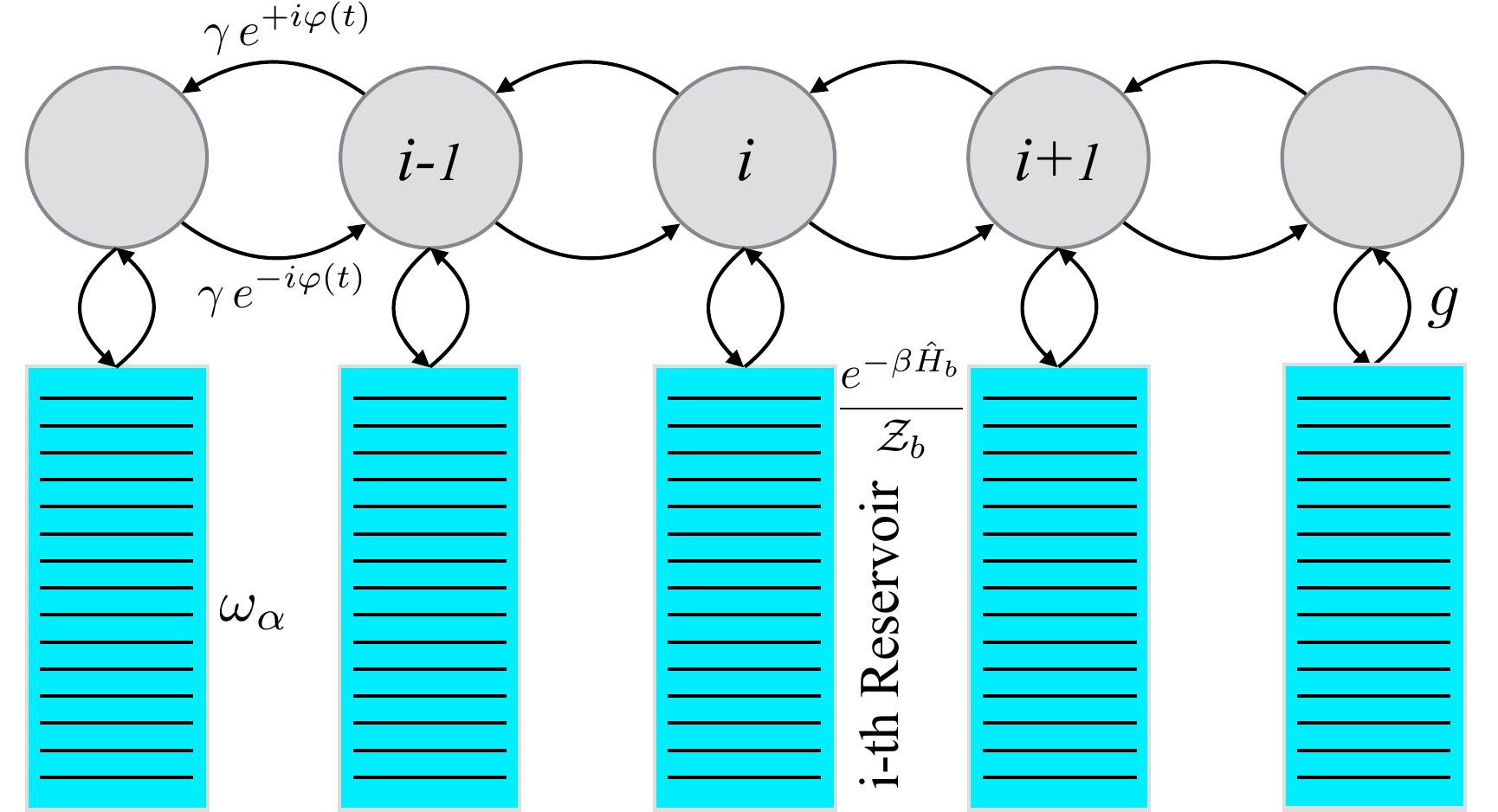}
\caption{Schematic representation of a one-dimensional tight-binding model, where electrons hop between nearest-neighbor sites under the effect of an electric field. Each lattice site is coupled to an infinite fermionic bath in thermal equilibrium that exchanges energy and fermions with the chain.}
\label{Fig:Ex}
\end{figure}
 In Fig.~\ref{Fig:Ex}, we schematically represent the linear chain coupled to the reservoir.
 The entire Hamiltonian of the system plus the bath is given by $\hat{\mathcal H}_{tot} = \hat{\mathcal H} +\hat{\mathcal H}_{b} +\hat{\mathcal V} $ and can be recast in a block diagonal form by expressing the fields in the Fourier basis, that is $d_k = \frac{1}{\sqrt{N}}\sum_n d_n e^{-ikn}$, $\cc_{k\alpha} = \frac{1}{\sqrt{N}}\sum_n \cc_{n\alpha} e^{-ikn}$ and their Hermitian conjugates. In this basis, the Hamiltonian decomposes into a sum of Hamiltonians for each momenta. 
In Ref.~\onlinecite{Han2012}, the dynamics of the electrons in the chain ($d$ fermions) are exactly solved by determining the nonequilibrium Green's function on the Keldysh contour; the problem can be solved exactly because it is quadratic in the fermion operators. 
\par In this section, we will introduce the master equation (ME) that defines the model that we will simulate on the quantum computer. In particular, we will derive the RME and subsequently a Lindbladian Master Equation (LME), which is easier to translate on a quantum circuit, showing what approximations we have to perform to obtain it.  In passing, we also compare the expected theoretical results with those of Ref.~\onlinecite{Han2012}, to show in which regimes our scheme significantly deviates from the physics that we want to simulate.  
\subsection{The  Master Equation}

\par The master equation governs the dynamics of the system's reduced density matrix $\hat{\rho} = \mbox{Tr}_b\, \hat{\rho}_{tot}$, where Tr$_b$ indicates the partial trace over the bath subspace. Within the Born approximation\cite{wiseman_milburn_2009}, the density matrix of the whole system is given by $\hat{\rho}_{tot} = \hat{\rho}\otimes\hat{\rho}_b(0)$, and we choose $\hat{\rho}_b(0) = \exp(-\beta\hat{H}_b)/\mathcal{Z}_b$, where $\beta$ is the inverse temperature of the bath. The $0$ argument on the bath density matrix denotes the initial start of the system at time $t=0$.

Given the block diagonal form of the full Hamiltonian (system plus bath), the system's reduced density matrix factorizes as a tensor product in momentum space, {\it i.~e.} $\hat{\rho} = \bigotimes_k \hat{\rho}^{(k)}$, meaning that we can define a $k$-dependent master equation for each $2\times 2$ $k$-dependent density matrix $\hat{\rho}_k$. 
 The master equation for each momentum subblock is:
%\begin{eqnarray}\label{master:gen}
%\partial_t{\hat{\rho}}_k+i[\hat{H}^{(k)},\hat{\rho}^{(k)}] &=& 
 %\hat{\mathcal{D}}^{(p)\dag}_{k}\,\hat{\rho}_k\,d_{k} -d_{k}
 %\hat{\mathcal{D}}^{(p)\dag}_k\,\hat{\rho}_k
% \nonumber \\
%&+& 
 %\hat{\mathcal{D}}^{(h)}_{k}\,\hat{\rho}_k\,d^{\dag}_{k} -d^
 %\dag_{k}\hat{\mathcal{D}}_{k}^{(h)}\,\hat{\rho}_k
%\nonumber \\
%&+& \mbox{h.~c.},
%\end{eqnarray}

\begin{eqnarray}\label{master:gen}
\partial_t\hat{\rho}_k&=&\mbox{Re}\,a^{\phantom{\dagger}}_k(t)\left[2\dd_k\hat{\rho}^{\phantom{\dagger}}_k d^{\phantom{\dagger}}_k-\left\{d^{\phantom{\dagger}}_k\dd_k,\hat{\rho}^{\phantom{\dagger}}_k \right\}\right] \nonumber \\
&+&\mbox{Re}A^{\phantom{\dagger}}_k(t)\left[2d_k\hat{\rho}^{\phantom{\dagger}}_k \dd_k-\left\{\dd_k d^{\phantom{\dagger}}_k,\hat{\rho}^{\phantom{\dagger}}_k \right\}\right],
\end{eqnarray}
where:
\begin{eqnarray}
\label{coeffs}
a_k(t) &=& g^2 \exp\left[-if_k(t) \right]\int_{-\infty}^0dt_1 \mathcal{C}_p(-t_1) \exp\left[if_k(t+t_1) \right], \nonumber \\
A_k(t) &=&  g^2 \exp\left[if_k(t) \right]\int_{-\infty}^0dt_1 \mathcal{C}_h(-t_1) \exp\left[-if_k(t+t_1) \right], \nonumber \\
%\label{eq:coeffs}
\end{eqnarray}
with $f_k(t) = \sin(k+\Omega t)/\Omega$, and where $\mathcal{C}_p(t)$ and $\mathcal{C}_h(t)$ are respectively the greater and lesser Green function of the bath fermions (see App.~\ref{appendix} for their definition and a derivation of these equations). In our case, we choose to attach an infinite bath to every site (see Fig.\ref{Fig:Ex}) that is at half-filling. In this, situation $\mathcal{C}_p(t) = \mathcal{C}_h(t)$.  Given the simple form of the bath Hamiltonian, the correlation function is $k$-independent and can be calculated analytically. In particular, in the limit of an infinite bandwidth with a flat density of states   [$N(\epsilon) \equiv \sum_\alpha \delta(\epsilon-\omega_\alpha)\sim N(0)$ (see Appendix \ref{appendix}, for further discussions about this limit)], we find that:
\begin{equation}\label{corr:func}
\mathcal{C}_p(t) =\pi N(0) \left[\delta(t)-\frac{i}{\beta}\mbox{PV}\,\mbox{cosech}\left(\frac{\pi t}{\beta}\right)\right].
\end{equation}
Here, PV denotes the principal value.

\par The coefficients that appear in Eq.~(\ref{coeffs}) are not always positive. This is made apparent by expanding the coefficients in terms of Bessel functions, as we  show in App.~\ref{appendix}. The loss of positivity can complicate our ultimate goal, i.e. to write a quantum circuit that simulates the driven-dissipative dynamics. In order to obtain a master equation that preserves positivity, we have to perform some additional approximations. In particular, we consider the following expansion:
\begin{equation}\label{eq:expansion}
f_k(t+t_1)\sim f_k(t) + \epsilon_k(t) t_1,
\end{equation}
in Eq.(\ref{coeffs}), which yields the following approximate coefficients:
\begin{eqnarray}
\label{coeffs:appr}
\mbox{Re}\,a_k(t) &\sim& \Gamma\, n_F\left[\epsilon_k(t)\right] \nonumber \\
\mbox{Re}A_k(t) &\sim&   \Gamma\, n_F\left[-\epsilon_k(t)\right].
\end{eqnarray}
The expansion that we consider is valid as long as the $t_1$ in the integral in Eq.~(\ref{coeffs}) can be considered small, \textit{i.~e.}~the time scale set by the correlation function of the bath, roughly $\tau_b = \beta/\pi$, is small compared to the period of  Bloch oscillations $\tau_b \ll  2\pi/\Omega$ and the inverse of the bandwidth $ \tau_b \ll 1/\gamma$. In this regard, we can consider the expansion in Eq.~(\ref{eq:expansion}) as perturbative in $\Omega$ and it is reminiscent of a recent approximation scheme to derive a Markovian time-dependent ME \cite{Dann2018}.  
 Eq.(\ref{eq:expansion}) 
 We shall refer to Eq.~(\ref{master:gen}), with the coefficients given in Eq.~(\ref{coeffs}), as the RME;  it is the LME when the coefficients are given in  Eq.~(\ref{coeffs:appr}).
 \subsection{Comparison of the RME, LME and exact solution}
 From Eq.~(\ref{master:gen}), we can obtain a differential equation for the momentum distribution function $n_k(t) = \mbox{Tr}\left(\hat{\rho}^{\phantom{\dagger}}_k(t) \dd_k\,d^{\,}_k\right)$, which can be immediately integrated to yield its solution. This is given by
\begin{equation}\label{master:nk}
\dot{n}_k  =  - 2\Gamma\, n_k + 2\mbox{Re}\,a_k(t).
\end{equation}

\begin{figure}
\includegraphics[width = \columnwidth]{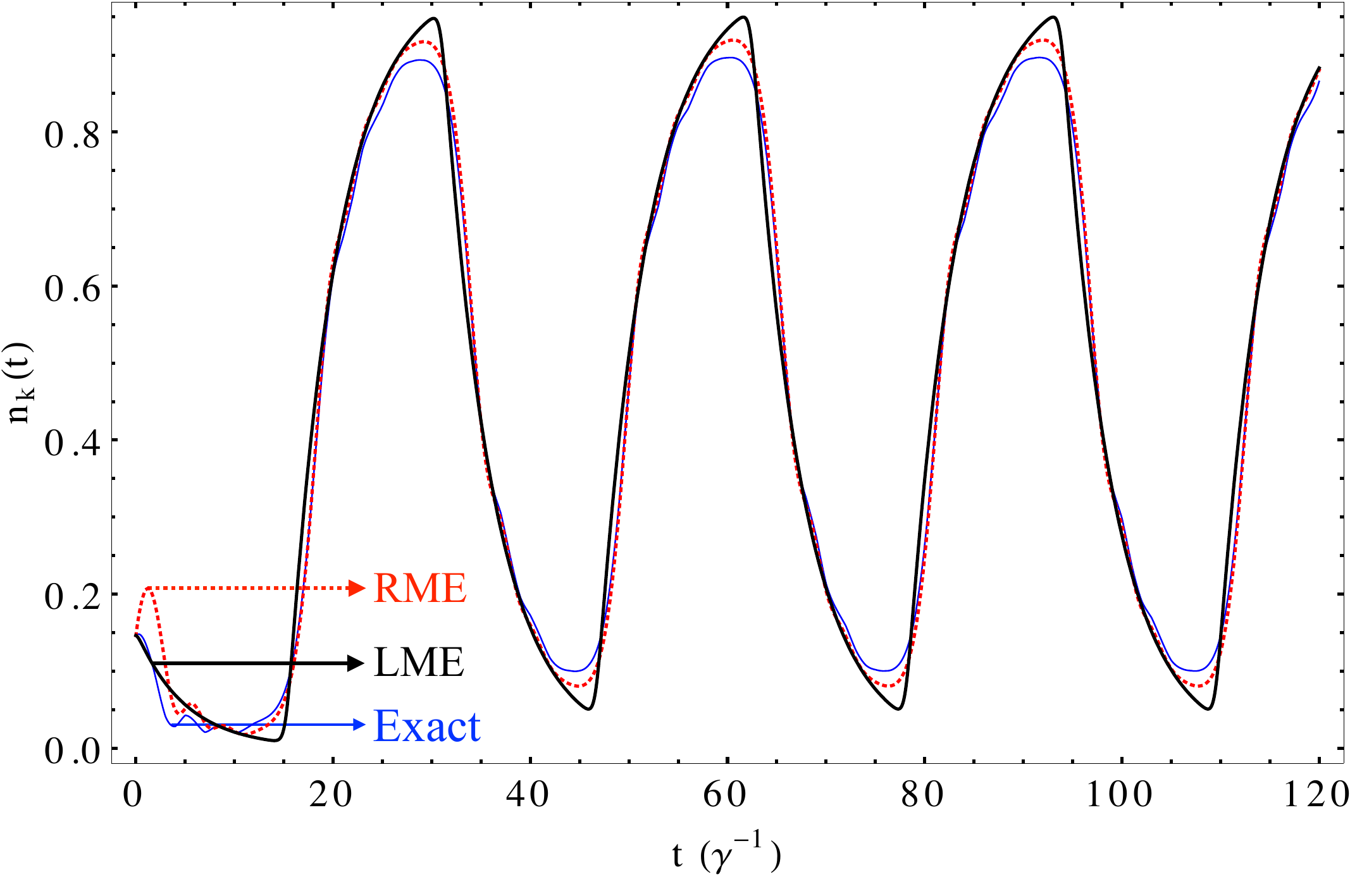}
\caption{
Momentum distribution function $n_k(t)$ as a function of time for 
$k = \pi/2+0.1$, 
$\Gamma/\gamma = 0.1$, and
$\Omega/\gamma = 0.2$.
We observe, that after an initial transient behavior, the occupation number reaches steady oscillations. We compare our results obtained by integrating the RME and LME, with the exact calculation\cite{Han2012} of $n_k(t)$.  See App.~\ref{appendix} for technical details.
}
\label{fig:mom:t}
\end{figure}

\par
\begin{figure}
\includegraphics[width = \columnwidth]{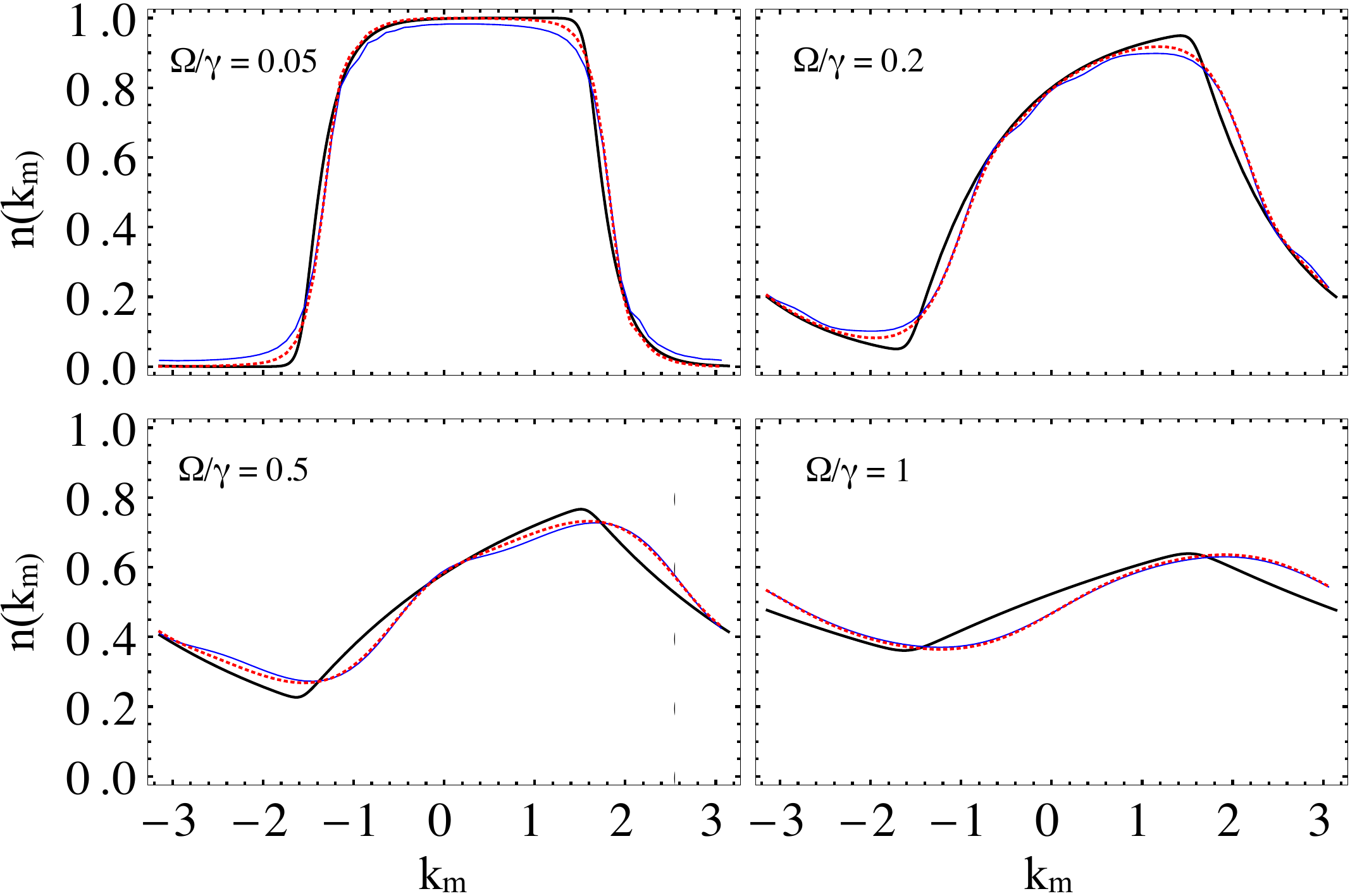}
\caption{Momentum distribution function $n(k_m)$ as a function of the gauge-invariant wavevector $k_m = k + \Omega t$ in the long time limit, for different values of $\Omega$ and for $\Gamma/\gamma = 0.1$. We compare the RME  (red dashed line) and the LME (black solid line) results with the exact solution (blue thin line) given in Ref.~\onlinecite{Han2012}. See App. \ref{appendix} for technical details.
}
\label{fig:mom:dist}
\end{figure}

 In Fig.~(\ref{fig:mom:t}), we show the time evolution of the momentum distribution function evaluated at $\Omega/\gamma = 0.2$, $\Gamma/\gamma = 0.1$, and $(k\,a) = \pi/2 + 0.1$. As one might expect for a driven-dissipative system, the system initially has transient behavior, which then evolves into steady oscillations at long times. The oscillatory behavior arises from the time-dependence of $a_k(t)$ [see Eq.~(\ref{mom:dist1})].
In Fig.~\ref{fig:mom:t}, we also plot the result obtained through the master-equation formalism. Comparing with the exact solution provided in Ref.~\onlinecite{Han2012}, we see excellent quantitative agreement, especially for long times. 
\begin{comment}
\par 
In the limit of $t_0\to-\infty$, Eq.~(\ref{mom:dist2}) becomes
\begin{equation}\label{mom:dist}
n_k(t) = 2\,\Gamma\,\mbox{Re}\,\sum_{\ell\ell^\prime} \frac{J_\ell\left(\frac{2\gamma}{\Omega}\right)J_{\ell^\prime}\left(\frac{2\gamma}{\Omega}\right)\mathcal{F}(\Omega \ell)}{2\Gamma-i(\ell-\ell^\prime)\Omega}e^{-i(\ell-\ell^\prime)(k + \Omega\, t)}
\end{equation}
\end{comment}
To understand the oscillating behavior at long times, we compute the occupation number as a function of the gauge-invariant wave vector $k_m = k + \Omega t$. 
Fig.~(\ref{fig:mom:dist}), shows $n(k_m)$ [from Eq.~(\ref{mom:dist})], where we replaced $k_m$ with $k + \Omega t$, as done in Ref.~\onlinecite{Han2012}, for different values of the driving field $\Omega$ at $\Gamma/\gamma = 0.1$. 
The momentum distribution shifts toward the driving field direction when $\Omega$ is increased, as expected since the electric field drives electrons in the direction of the field. When the field is large, the momentum distribution function loses its original shape and becomes sinusoidal with a smaller width (this is probably due to a tendency toward a Wannier-Stark ladder\cite{wannier}, but with broadening due to the dissipation). We compare our results with the exact solution and find excellent quantitative agreement between the RME and the exact solution for all values of the field strength.
The LME predictions deviate from the exact solution when $\Omega$ is sufficiently large, as expected given the perturbative nature of the expansion in Eq.~(\ref{eq:expansion}) that we used to set the LME. We observe that for large field values the $n(k_m)$ profile obtained with the LME is shifted by a phase compared to the exact solution and its shape tends to a triangular wave rather than a sinusoidal. However, the amplitude and mean value of the oscillations appear to agree with the exact solution also for large $\Omega$.

\par The total current of the system in the steady state is given by the formula $J = (2\pi)^{-1}\int dk_m\, v_{k_m} n_{k_m}$, where the band velocity is defined as $v_k = \partial_k \epsilon_k$, with the band structure given by $\epsilon_k = -2\gamma\,\cos k$. For generic times the current is a time-dependent function, however the long-time behavior of $J(t)$ is a constant in time. It is evident that in order to obtain a finite current, $n_{k_m}$ must not be symmetric with respect to the origin (not an even function), and we have already shown that a finite electric field tends to distort the momentum-distribution function towards the field direction [see Fig.~\ref{fig:mom:dist}]. 
In App.~\ref{appendix}, we give the analytic expressions for the current in the long-time limit for both the RME and the LME, here we will limit ourselves to show and comment on the results. 
\begin{comment}
\begin{eqnarray}\label{current}
\langle J\rangle &=& 4\gamma\,\Gamma\,\mbox{Re}\sum_\ell 
\frac{J_\ell\left(\frac{2\gamma}{\Omega}\right)
J_{\ell+1}\left(\frac{2\gamma}{\Omega}\right)
\mathcal{F}(\Omega \ell)}{\Omega-2i\Gamma} \nonumber \\
&+&4\gamma\,\Gamma\,\mbox{Re}\sum_\ell 
\frac{J_\ell\left(\frac{2\gamma}{\Omega}\right)
J_{\ell-1}\left(\frac{2\gamma}{\Omega}\right)
\mathcal{F}(\Omega \ell)}{\Omega+2i\Gamma} \,.
\end{eqnarray}
\end{comment}

\begin{figure}
\includegraphics[width= \columnwidth]{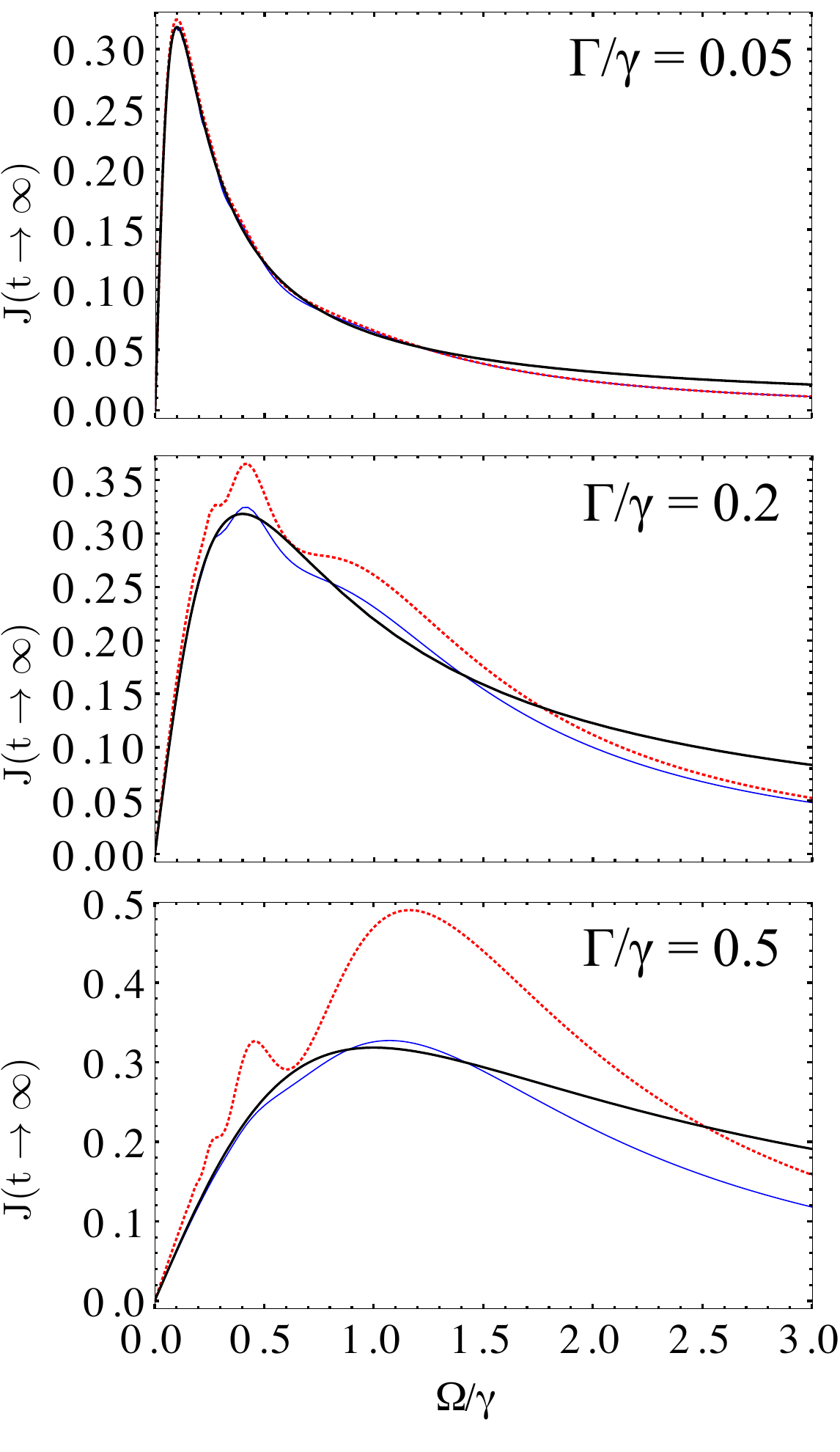}
\caption{DC current in the long-time limit as a function of the driving field for different values of the coupling with the bath $\Gamma$. Our results from the RME (red dashed lines) and LME (black solid line) are compared with the exact curves of the current (blue thin lines) calculated in Ref.~\onlinecite{Han2012}. See App. \ref{appendix} for technical details.
}\label{fig:curr}
\end{figure}

The dc current as a function of the electric field is shown in Fig.~\ref{fig:curr}. For small enough values of $\Omega$, the current grows linearly with the electric field, illustrating the expected Ohm's law-behavior in the linear-response regime. When the field intensity increases and becomes comparable to the dissipation rate $\Gamma$, a Bloch electron has enough life time to reach the Brillouin zone boundary.
In this regime, Bloch oscillations effects become important and the dc current first reaches a maximum at $\Omega \sim 2 \,\Gamma$. When $\Omega$ is further increased, the period of the oscillations is very high so that Bloch electrons reach the BZ boundaries and are reflected back multiple times, decreasing the expectation value of the current, that averages to zero in the limit $\Omega \to \infty$.

We compare our result with the exact one calculated in Ref.~\onlinecite{Han2012}. The agreement is excellent for small values of $\Gamma$ for both the RME and LME results. Note that when the coupling with the bath is increased, the master-equation prediction starts to deviate from the exact result. However, as shown in Fig.~\ref{fig:curr} for $\Gamma/\gamma = 0.2$, there is a region of intermediate values of $\Gamma$ where the RME reproduces all the qualitative features of the exact solution; it just has some quantitative mismatch. This disagreement is largest for intermediate values of $\Omega$. 
When the dissipation rate is increased further the RME stops to be a reliable approximation, as we can see in Fig.~\ref{fig:curr} for $\Gamma/\gamma = 0.5$, where also in this case the  disagreement is maximal for intermediate values of $\Omega$. 
Instead, in the asymptotic limit when $\Omega \to \infty$, the RME seems to give good results also for large values of $\Gamma$.
It is worthwhile to note that the LME predictions are in an excellent quantitative agreement with the exact solution for a broad range of $\Omega$ and $\Gamma$. This is surprising given that the LME was obtained in the limit of both $\Gamma, \Omega \ll \gamma$.  We note that only in the asymptotic limit do we start to appreciate a mismatch  between the exact solution and the LME, which is enhanced by increasing $\Gamma$.
This might be a peculiarity of our model and further investigations are needed to establish if the LME yields good results in the strongly driven regime when  on-site Coulomb   repulsion between electrons is taken into account. However, this would go beyond the aim of our paper and we postpone it to future study.
\section{The Kraus Maps}\label{sec:III}

Because we are interested in the possibility of simulating dissipative dynamics on a quantum computer, we now explore some of the details behind how one would do this. We devise two schemes:
\begin{enumerate}
    \item[(i)] The dynamics of the system is solved directly by the quantum computer using Trotter steps;
    \item[(ii)] We show how to dissipatively prepare the long-time NESS in one step. 
\end{enumerate}
To create a better connection with quantum computation, it is useful to express the time evolution of the density matrix in terms of the operator-sum representation:  
\begin{equation}\label{def:sum}
    \rho(t) = \sum_i K_i(t)^{\,} \rho(t_0)K_i^{\dag}(t), 
\end{equation}
where $\rho(t)$ is the system density matrix at time $t$, $\rho(t_0)$ is set by the initial condition, and $K_i$ are the Kraus operators that satisfy the following sum rule:
\begin{equation}\label{sum:rule}
    \sum_i K^\dag_i K^{\,}_i = \mathbbm{1}.
 \end{equation}
 In our case, the set of $K_i$ depends on the master equation we started with and on the particular scheme that we want to adopt: (i) or (ii). 
 \subsection{The Trotterised map}\label{sec:trotterised}
 In general, determining the map in Eq.~(\ref{def:sum}) is tantamount to finding the exact solution of the problem, which would be very challenging especially for large enough systems.
Instead, it is possible to construct an infinitesimal map that evolves a state from an initial time $t$ to a final state at time $t + d t$, when the master equation is  in Lindblad form \cite{benenti2007principles}. In fact in this case, the Kraus map is given by $K_1 = \mathbbm{1} - iH dt -\frac{1}{2}\sum_{i>1}L_i^\dag L_i^{\,} dt$, and $K_{i>1} = \sqrt{d t}L_i$, where $L_i$ are the Lindbladians, and $\sum_{i = 1}^4 K_i^{\dag}K_i^{\,} = \mathbbm{1} + O(dt^2)$. \par In analogy with  algorithms for closed systems \cite{Lloyd1996}, where the dynamics of a quantum state are obtained in an approximate fashion using Trotter steps, we can construct  a semi-positive trace preserving map, that satisfies exactly the sum rule in Eq.~(\ref{sum:rule}), and that reconstructs approximately the state at time $t + \Delta t$ from the state at time $t$, performing the following mapping:
\begin{equation}\label{trotter:mapping}
\rho(t)\mapsto\rho(t+\Delta t)=\sum_i K_i(t)\rho(t) K_i^\dagger(t),
\end{equation}
where $\Delta t$ is finite and:
\begin{eqnarray}\label{Trotter:Map}
K_1 &=&\exp \left( i H(t) \Delta t \right)\sqrt{\mathbbm{1}-\sum_i L^\dag_{i}(t) L^{\,}_i(t)\Delta t }  \nonumber \\
K_{i}&=&\sqrt{\Delta t}L_i(t) \hspace{1cm}\mbox{when }i>1. 
\end{eqnarray}
We note that the map defined by  Eqs.~(\ref{trotter:mapping},\ref{Trotter:Map}) recovers the infinitesimal map when expanded to first order in $\Delta t$, i.e. $K_1 \sim \mathbbm{1}-iH(t) \Delta t-\frac{1}{2}\sum_i L^\dag_i L_i^{\,}\Delta t$, and when the dissipation rates go to zero, it gives back the time evolution of the isolated system, \textit{i.~e.}~$K_1 = \exp(i H \Delta t)$ and $K_{i>1} = 0$. We note further that such a construction is generic and can be applied to any system when the Hamiltonian and the Lindbladians are specified. 
\par  Let us now consider as an example our system of non-interacting electrons in an electric field. 
In this case, we might be tempted to define as Lindbladians the following operators $\sqrt{2 \mbox{Re}a_k(t)}d^{\,}_k$ and $\sqrt{2 \mbox{Re}A_k(t)}\dd_k$. However, this would not correspond to a trace preserving map, when Re$\,a_k(t)$ or Re$A_k(t)$ become negative. Therefore, this scheme is suitable for simulating only the LME where the time-dependent coefficients are defined in Eq.~(\ref{coeffs:appr}). 
\par Under these assumptions, the Trotterised Kraus map becomes:
\begin{eqnarray}\label{map:trot:LME}
K_1 &=& \sqrt{1-2\, \Gamma \,n_F[-\epsilon_k(t)]\Delta t}\, P_1+ \sqrt{1-2\, \Gamma \,n_F[\epsilon_k(t)]\Delta t}\,P_0\nonumber \\
K_2 &=&\sqrt{2\,\Gamma\,n_F[-\epsilon_k(t)] \Delta t}\, X P_1\nonumber \\
K_3 &=& \sqrt{2\,\Gamma\,n_F[\epsilon_k(t)] \Delta t}\, X P_0,
 \end{eqnarray}
 where we defined $P_0 = d^{\,}_k d^{\dag}_k$, $P_1 = d^{\dag}_k d^{\,}_k$ and $X = d^\dag_k + d^{\,}_k$.
 We note that Eq.~(\ref{map:trot:LME}) gives a constraint to the maximum allowed time step, which is given by $\Delta t < 1/2\Gamma$. 
\subsection{The integrated map}\label{sec:int:map}
In this section, we determine the integrated Kraus map for our system of non-interacting electrons. We closely follow the work by Andersson \textit{et al.}~\cite{Andersson2007}.
\par We first express the density matrix at time $t$ [$\rho_k(t)$ from Eq.~(\ref{master:gen})] as a map from its initial value $\rho_k(0)$ via
\begin{equation}
\rho_k(t) = \phi_t(\rho_k(0)) = \sum_{a=0}^3\sum_{b=0}^3S_{ab}(t)\sigma_a\rho_k(0) \sigma_b,
\end{equation}
where $\sigma_0 = \mathbbm{1}_{2\times 2}/\sqrt{2}$, $\sigma_1 = \sigma^x/\sqrt{2}$, $\sigma_2 = \sigma^y/\sqrt{2}$, $\sigma_3 = \sigma^z/\sqrt{2}$, with $\sigma^{\alpha = \{x,y,z\}}$ being the standard $2\times 2$ Pauli matrices and where $S_{ab}(t)$ is the so-called Choi matrix (which is a Hermitian and time-dependent $4\times 4$ matrix) expressed in the Pauli basis (the time evolution of the density matrix is a positive trace-preserving map). $\phi_t$ and $S_{ab}$ generally have a $k$ dependence which we have omitted for readability. Note that we are working in a specific fixed momentum subspace, so the density matrix here is a $2\times 2$ matrix and the Choi matrix is a $4\times 4$ matrix; the indices in the summations run over only four values. We can rewrite this map in a diagonalized form in the following way:
\begin{equation}\label{K:map}
\phi_t(\rho_k(0)) = \sum_{i=1}^4 K_i(t)\rho_k(0) K^\dag_i(t),
\end{equation}
where $K_i(t) = \sqrt{\lambda_i}\sum_{a=0}^3 X(i)_a \sigma_a$ are the Kraus operators with $\lambda_i$ and $X(i)$ being the eigenvalues and eigenvectors of the Choi matrix (the Kraus operators are effectively square-roots of the Choi matrix). Note that the Kraus operators are expressed as operators in a two-dimensional space, given by the Pauli matrices.

\par We can obtain the Choi matrix directly from the master equation by realizing that the equation of motion can be re-expressed as $\partial_t\rho_k(t) = \Lambda_t(\rho_k)$, where $\Lambda_t$ is a linear map such that $\Lambda_t(\rho_k)$ is Hermitian and traceless. Using this form, it can be shown that\cite{Andersson2007} \begin{equation}
S_{ab}(t) = \sum_{r=0}^3\sum_{s=0}^3F_{sr}(t)\mbox{Tr}\left[\sigma_r\sigma_a\sigma_s\sigma_b\right],
\end{equation}
where $F_{rs} = \mbox{Tr}(\sigma_r\phi_t(\sigma_s))$ is a matrix representation of the linear map $\phi_t$. In App.~\ref{linear:map}, we give more details on the analytical derivation of the linear map $\phi_t$. This matrix is related through a differential equation to the matrix representation of $\Lambda_t$ in the following way: $\dot{F}(t) = L(t)\cdot F(t)$, where "$\cdot$" indicates the matrix product, with initial condition $F(0) = \mathbbm{1}_{4\times 4}$ and with $L_{rs} = \mbox{Tr}(\sigma_r\Lambda_t(\sigma_s))$.

\par In general, finding the Choi matrix is tantamount to finding an exact solution of the problem, so it can be quite complex. But, for the noninteracting system that we consider here, the procedure is greatly streamlined. Indeed, an analytic solution is possible, as we now show. For long enough times, the Kraus operators become
%$K_1 = \sqrt{1-n_k(t)}\ket{0}\bra{0}$, $K_2 = \sqrt{n_k(t)}\ket{1}\bra{1}$, $K_3 = \sqrt{n_k(t)}\ket{0}\bra{1}$, and $K_4 = \sqrt{1-n_k(t)} \ket{1}\bra{0}$, 
\begin{equation}\label{kraus_map}
\begin{split}
K_1 &= \sqrt{1-n_k(t)}\,P_0\\
K_2 &= \sqrt{n_k(t)}\,P_1\\
K_3 &= \sqrt{n_k(t)}\,X\, P_0\\
K_4 &= \sqrt{1-n_k(t)}\, X\,P_1\\
\end{split}
\end{equation}
where $n_k(t)$ is defined in Eq.~(\ref{mom:dist}), and we generically focus on taking the long-time limit. Note how we need to know the final momentum distribution in order to determine these Kraus operators, again indicating that determining them is equivalent to completely solving the system. Given fermionic statistics, namely that $0\le n_k(t)\le 1$, these long-time Kraus operators satisfy the normalization condition $\sum_i K^\dag_iK_i = \mathbbm{1}_{2\times 2}$. When the Kraus map is applied to a generic initial state $\rho_k(0)$, it returns the following time-dependent mixed state:
\begin{equation}\label{final_state}
    \rho_k(t) = \phi_t(\rho_k(0)) = [1-n_k(t)]\ket{0}\bra{0} + n_k(t)\ket{1}\bra{1}
\end{equation}
which depends only on the population of electrons with momentum $k$. We call it the steady-state density matrix or $\rho_{\rm ss}$. 

\section{Quantum Circuits}\label{sec:IV}
\subsection{Quantum Simulation of Driven-Dissipative Dynamics}
\par Now we discuss how one can simulate the action of the Kraus map given in Eq.~(\ref{map:trot:LME}) by using a quantum circuit. The circuit implementing a single Trotter step must perform the mapping written in Eq.~(\ref{trotter:mapping}).
%We imagine that the system has been prepared in some arbitrary initial state $\rho(0)$. Then the circuit implementing a single Trotter step must perform the following mapping

This is not a unitary map (because we are simulating dissipation), so ancilla qubits must be employed to purify the channel into a unitary operation. Such a unitary operator is guaranteed to exist by Stinespring's dilation theorem\cite{preskill_notes}. 
%Taking a partial trace over the ancilla recovers the reduced density matrix of the system. 

We may interpret this Kraus map as performing the following: If the system is in state $\ket{0}$, apply $X$ with probability $k_1\equiv2\Gamma n_F[\epsilon_k(t)]\Delta t$ and apply the identity  $I$ with probability $1-k_1$. If the system is in state $\ket{1}$, apply $X$ with probability $k_2\equiv2\Gamma n_F[-\epsilon_k(t)]\Delta t$ and apply $I$ with probability $1-k_2$.

\begin{figure}[H]
	\includegraphics[width=\columnwidth]{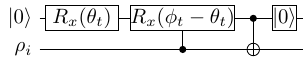}
	\caption{Circuit fragment inducing correct transitions in the system. The $\ket{0}$ gate indicates a selective reset of that qubit to the $\ket{0}$ state. $\theta_t=2\arcsin\sqrt{2\Gamma n_F[\epsilon_k(t)]\Delta t}$ and $\phi_t=2\arcsin\sqrt{2\Gamma n_F[-\epsilon_k(t)]\Delta t}$. This will give the correct diagonal terms in the resulting density matrix.}\label{fig:trot_circ2}
\end{figure}

One way of doing this is as shown in Fig.~\ref{fig:trot_circ2}. Here we have defined $\theta_t=2\arcsin\left(\sqrt{k_1} \right)$ and $\phi_t=2\arcsin\left(\sqrt{k_2} \right)$. We begin by rotating the ancilla qubit by $\theta_t$, then if the system is in $\ket{0}$, we do nothing, and if the system is in $\ket{1}$, we undo the rotation of $\theta_t$ and rotate by $\phi_t$. This leaves the ancilla in $\ket{1}$ with probability $P(1|0)=\sin^2{\theta_t/2}=k_1$ and $P(1|1)=\sin^2{\phi_t/2}=k_2$, as desired. Finally, we flip the system qubit if the ancilla is $\ket{1}$. This accomplishes the operation described above.

However this is not sufficient to implement the proper Kraus map. The form of the Kraus Map arises from taking the partial trace over the ancillary degrees of freedom of the system+ancilla density matrix, after evolving with the joint time evolution operator. Calling the initial state of the ancilla $\ket{a_0}\bra{a_0}$ and choosing $\{\ket{a_i}\}$ as the basis for the ancilla, this gives
\begin{equation}
\rho(t)=\sum_i \underbrace{\bra{a_i}U(t)\ket{a_0}}_{K_i(t)}\rho(0)\underbrace{\bra{a_0}U^\dagger(t)\ket{a_i}}_{K_i^\dagger(t)}.
\end{equation}
This means that to get the proper system density matrix after tracing over the ancilla, the circuit must induce the desired transitions, $\{K_i(t)\}$, on the system qubit \textit{as well as} map $\ket{a_0}\mapsto\ket{a_i}$ in the ancilla register for the corresponding $K_i$. That is, we seek a circuit implementing a unitary $U$ that accomplishes the mapping given in Table \ref{tab:map1}.\\
\begin{table}
    \centering
    \begin{tabular}{c|c}
    Initial State  & Final State
    \\\hline
    $\ket{0}\ket{a_0}$ & $\sqrt{k_1}\ket{1}\ket{a_1}+\sqrt{1-k_1}\ket{0}\ket{a_0}$\\
    $\ket{1}\ket{a_0}$ &$\sqrt{k_2}\ket{0}\ket{a_2}+\sqrt{1-k_2}\ket{1}\ket{a_0}$
    \end{tabular}
    \caption{Action of Trotterized Kraus Map. $\{\ket{a_i}\}$ are any three orthogonal states in the ancilla register, $k_1=2\Gamma n_F[-\epsilon_k(t)]\Delta t$, and $k_2=2\Gamma n_F[\epsilon_k(t)]\Delta t$.}
    \label{tab:map1}
\end{table}

\begin{figure}[H]
	\includegraphics[width=\columnwidth]{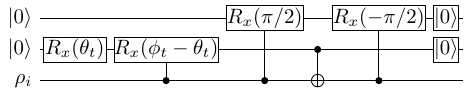}
	\caption{Circuit implementing a single Trotter step of the Kraus map given in Eq.~\ref{map:trot:LME}. The $\ket{0}$ gate indicates a selective reset of that qubit to the $\ket{0}$ state. In addition, we have $\theta_t=2\arcsin\sqrt{2\Gamma n_F[\epsilon_k(t)]\Delta t}$ and $\phi_t=2\arcsin\sqrt{2\Gamma n_F[-\epsilon_k(t)]\Delta t}$. For the initial state $\rho_i=\rho(t)$, the final state of that same qubit is $\rho(t+\Delta t)$.
	%To perform the next trotter step, the ancilla register must be reset to $\ket{00}$, either by resetting the qubits or swapping them out for new qubits.
	}\label{fig:trot_circ3}
\end{figure}

The circuit shown in Fig. \ref{fig:trot_circ2} implements a $U$ which results in $\ket{a_1}=\ket{a_2}$; this gives incorrect off-diagonal terms in the density matrix. This is not an issue for this study, since our quantity of interest, $n_k(t)$, is given solely by the diagonal terms, whose evolution is not influenced by the off-diagonals. Furthermore, the steady-state density matrices produced by the circuit from Fig.~\ref{fig:trot_circ2} as well as from Eq.~\ref{map:trot:LME} are identical, being purely diagonal because our eigenstates are computational basis states. However, if one wishes to access the proper transient states---the circuit given in Fig.~\ref{fig:trot_circ3} implements a Trotter step of the Kraus map in Eq.~\ref{map:trot:LME} exactly. We verify the equivalence of the circuits and Kraus maps by recasting both in terms of matrix operations and directly comparing their action on a generic $\rho$.

\begin{figure}[h]
	\includegraphics[width=\columnwidth]{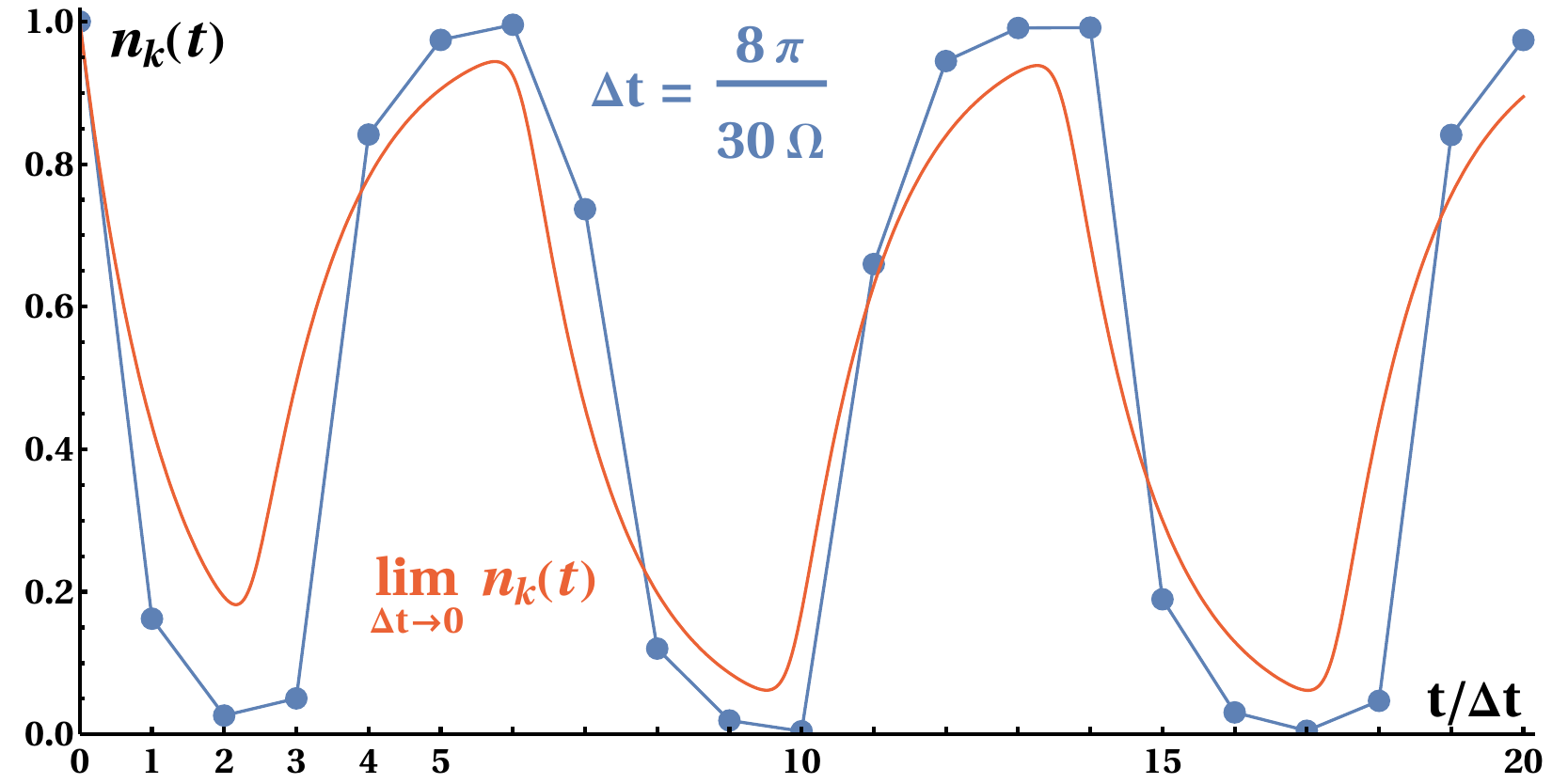}
	\caption{Comparison of numerically solving the Lindblad master equation from Eq.~(\ref{master:gen}) using the coefficients in Eq.~(\ref{coeffs:appr}) versus using a finite Trotter step size. We have used $\Gamma=0.1$, $\Omega=0.2$, $\beta=5$, $\Delta t=8\pi/30\Omega$, $k=7\pi/8$ and $\rho(0)=\ket{1}\bra{1}$}\label{fig:TrotVsEx}
\end{figure}

Fig.~\ref{fig:TrotVsEx} shows the result for $n_k(t)$ obtained by numerically solving the Lindblad master equation from Eq.~(\ref{master:gen}) using the coefficients in Eq.~(\ref{coeffs:appr}), plotted against $n_k(t)$ as obtained by classically iterating either circuit given in Figs.  \ref{fig:trot_circ2},\ref{fig:trot_circ3} (both circuits give identical results). Being a classical simulation of the circuit, the discrepancies between the two curves shown in Fig. \ref{fig:TrotVsEx} arise only from using a finite $\Delta t$. Indeed the two curves converge as $\Delta t\to0$. Ideally we would like to use a very small $\Delta t$ and take a large number of steps to avoid the error associated with finite $\Delta t$, however this is not possible on current hardware as discussed below.

We ran the circuits shown in Figs. \ref{fig:trot_circ2} and \ref{fig:trot_circ3} on IBMQ's quantum hardware\cite{Qiskit}. The resulting output was corrected by the ``pseudo-inverse" method as outlined in Qiskit's ``Measurement Error Mitigation" tutorial and implemented in Qiskit Ignis\cite{Qiskit}. The data from the circuit in Fig.~\ref{fig:trot_circ2} is in good agreement overall with the predicted behavior at early times, beginning to deviate for $t>4\Delta t$. The data from the circuit in Fig.~\ref{fig:trot_circ3} is expectedly worse at all times, and begins to show significant deviations for $t>3\Delta t$.

\begin{figure}[h]
	\includegraphics[width=\columnwidth]{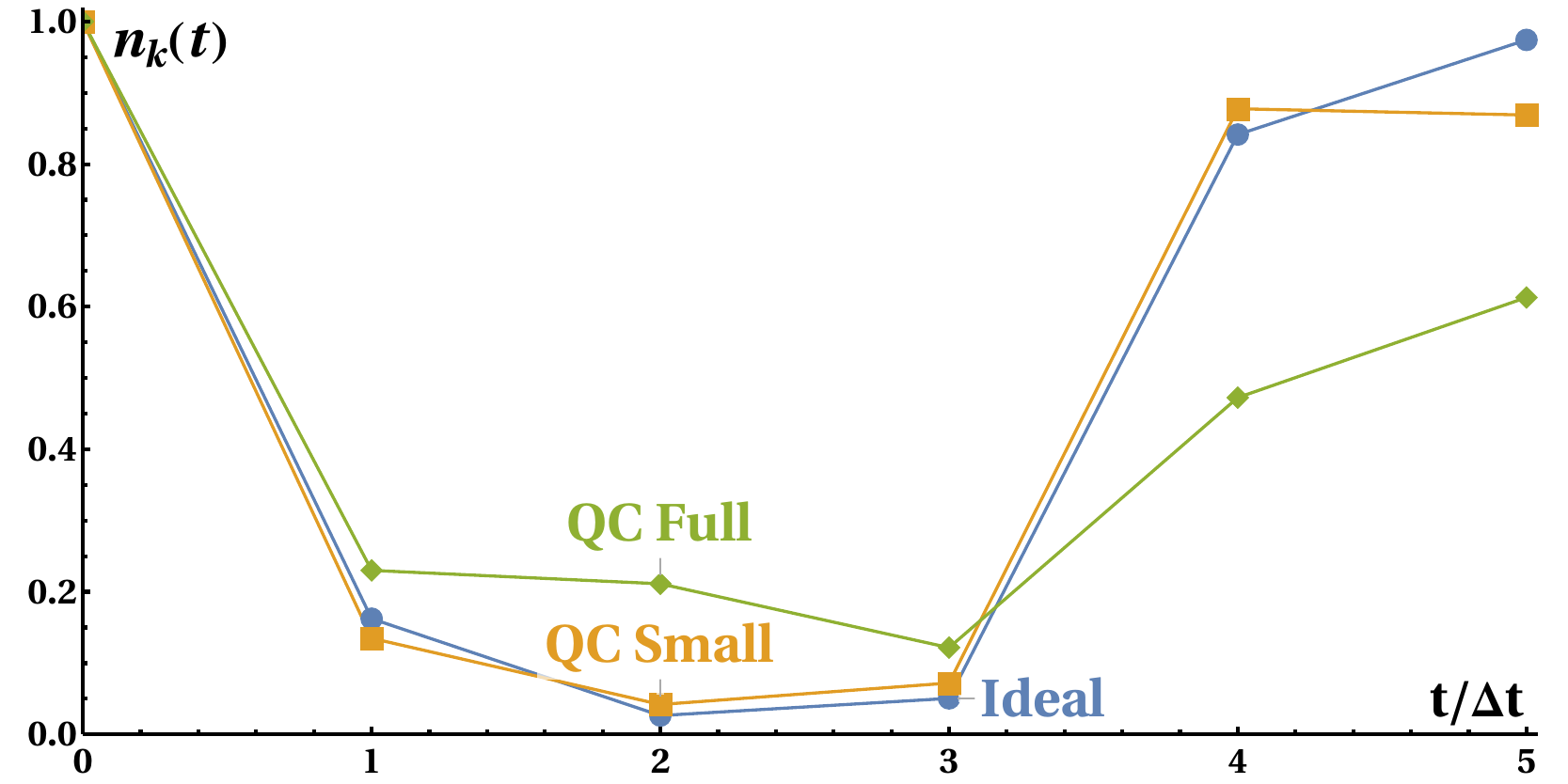}
	\caption{Result from iteratively running the circuits from Figs.~\ref{fig:trot_circ2} and \ref{fig:trot_circ3} (orange and green respectively) on IBMQ's Singapore machine. We have used $\Gamma=0.1$, $\Omega=0.2$, $\beta=5$, $\Delta t=8\pi/30\Omega$, $k=7\pi/8$ and $\rho(0)=\ket{1}\bra{1}$. The idealized results, from Fig.~\ref{fig:TrotVsEx}, are included in blue for reference.
	Note that the time axis runs only out to 5 units here because we can only implement a small number of Trotter steps.}\label{fig:TrotRes}
\end{figure}

As can be seen from Fig.~\ref{fig:TrotVsEx}, many Trotter steps must be run in order to reach the steady state. This is problematic since, at the time of this writing, most quantum hardware, including IBMQ, do not support selective reset of qubits. This means that each additional Trotter step requires swapping in additional fresh qubits in lieu of an actual reset. Given the limited connectivity of these devices, fidelity drops off quickly as more and more distant qubits are required to be swapped into position near the system qubit. Ultimately selective reset capabilities will be required to implement such protocols on near-term devices with limited connectivity and qubits. The results of these runs are shown in Fig.~\ref{fig:TrotRes} and the computational cost of each step is given in Table \ref{tab:trotCost}.

\begin{table*}
    \centering
    \begin{tabular}{l|c|c|c|c|c}
    Trotter Step & 1 & 2 & 3 & 4 & 5
    \\\hline
    \begin{tabular}{l}
    q.b. = qubits \\
    Fig. \ref{fig:trot_circ2} Circuit\\
    Fig. \ref{fig:trot_circ3} Circuit
    \end{tabular}
    &
    \begin{tabular}{ccc}
    $R_x$ & c$X$ & q.b. \\\hline
    3 & 3 & 2\\
    7 & 7 & 3
    \end{tabular}
    &
    \begin{tabular}{ccc}
    $R_x$ & c$X$ &  q.b. \\\hline
    6 & 6 & 3\\
    14 & 20 & 5
    \end{tabular}
    &
    \begin{tabular}{ccc}
    $R_x$ & c$X$ &  q.b. \\\hline
    9 & 9 & 4\\
    21 & 34 & 7
    \end{tabular}
    &
    \begin{tabular}{ccc}
    $R_x$ & c$X$ &  q.b. \\\hline
    12 & 15 & 5\\
    28 & 72 & 9
    \end{tabular}
    &
    \begin{tabular}{ccc}
    $R_x$ & c$X$ &  q.b. \\\hline
    15 & 21 & 6\\
    35 & 78 & 11
    \end{tabular}
    \end{tabular}
    \caption{Quantum resources for running circuits in Figs.~\ref{fig:trot_circ2}, \ref{fig:trot_circ3}. Each c$R_x$ is decomposed as 2 $R_x$ and 2 c$X$ gates. $R_x$ gates are implemented as IBM's native $U_3$ gate. Limited connectivity creates an accelerating c$X$ cost as ever more SWAP operations are needed to bring in fresh ancilla.}
    \label{tab:trotCost}
\end{table*}

\subsection{Dissipative Quantum State Preparation of the Non-equilibrium Steady State}

Due to the lack of selective reset capabilities, it is generally not possible to run our protocol far enough out in time to reach and investigate the steady-state dynamics of our system. Note that for Eq. \ref{map:trot:LME} to represent a physical map, we must have $\Delta t\leq 1/(2\Gamma)$, so we cannot take arbitrarily large steps even if we were willing to accept the associated error. Therefore we present a quantum circuit implementing the integrated Kraus map given in Eq.~\ref{kraus_map}. Given any initial state, pure or mixed, this circuit prepares the desired steady state in a single step and thus circumvents the need to employ partial reset gates. The obvious drawbacks here are that the steady state must be known in advance and one cannot access the transient dynamics.

Following the procedure given in the preceding section, we seek a circuit implementing a unitary $U$ that accomplishes the mapping given in Table \ref{tab:map2}. This is very similar to the map above with two key differences.

First, we have four distinct ancillary states, which allows us to replace the two c$R_x(\pm \pi)$ gates in the circuit in Fig.~\ref{fig:trot_circ3} with a single c$X$ gate. This is because in the Trotterized map, we need that an application of $I$ on the system, for \textit{both} possible system states, maps the ancilla to the same state; the integrated map, however, requires that $I$ applied on the system in $\ket{0}$ and $I$ applied to the system in $\ket{1}$ leaves the ancilla in distinct states, which is precisely what a c$X$ does.

\begin{table}
    \centering
    \begin{tabular}{c|c}
    Initial State  & Final State
    \\\hline
    $\ket{0}\ket{a_0}$ & $\sqrt{n_k(t)}\ket{1}\ket{a_1}+\sqrt{1-n_k(t)}\ket{0}\ket{a_0}$\\
    $\ket{1}\ket{a_0}$ &$\sqrt{1-n_k(t)}\ket{0}\ket{a_2}+\sqrt{n_k(t)}\ket{1}\ket{a_3}$
    \end{tabular}
    \caption{Action of integrated Kraus map. $\{\ket{a_i}\}$ are any four orthogonal states in the ancilla register, and $n_k(t)$ is given in Eq.~(\ref{mom:dist}).}
    \label{tab:map2}
\end{table}

\begin{figure}[hbtp!]
	\includegraphics[width=\columnwidth]{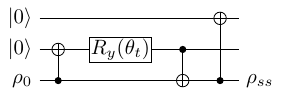}
	\caption{Final version of the circuit implementing the integrated Kraus map from Eq.~\ref{kraus_map}, which was run on IBM's Boeblingen quantum computer\cite{ibm}. Here we have $\theta_t=2\arcsin\sqrt{n_k(t)}$}
	\label{q:circuit3}
\end{figure}

Second, we have $k_1=1-k_2\equiv n_k(t)$. This means $\theta_t=\pi-\phi_t$ and this symmetry allows us to convert the $R_x(\theta_t)$ and c$R_x(\phi_t-\theta_t)$ into $R_y(\theta_t)$ and c$X$. Because of this, we can simplify the circuit in Fig.~\ref{fig:trot_circ3} giving the circuit in Fig.~\ref{q:circuit3}. We again verify the equivalence of the circuit and Kraus map by recasting both in terms of matrix operations and directly comparing their action on a generic $\rho$. A similar circuit to that one shown in Fig.(\ref{q:circuit3}) has been run on an IBM Quantum Experience platform to reproduce topological thermal states \cite{Viyuela2018}.
%We use the convention that qubit $1$  is the top line and qubit $3$ is the bottom (system) line. Defining c$X_i^j$ be the operation which applies $cX$ controlled on $i$, targeting $j$ and the identity on all other qubits, we find that our circuit is given by
%\begin{equation}
%U=
%(\text{c}X_3^1)
%\cdot 
%(\text{c}X_2^3)
%\cdot 
%[I\otimes I\otimes R_y(\theta)]
%\cdot
%(\text{c}X_3^2).
%\end{equation}
% This gives the final density matrix of the joint system+ancilla registers as
% \begin{equation}
% \rho_3=U\left[\ket{00}\bra{00}\otimes\left(
% \begin{array}{cc}
% a & b \\
% b^* & 1-a \\
% \end{array}
% \right) \right]U^\dagger,
% \end{equation}
% where $\rho(0)=\left(\begin{array}{cc}
% a & b \\
% b^* & 1-a \\
% \end{array}
% \right)$ is a generic input state. Taking the partial trace of $\rho_3$ over the ancilla returns the reduced density matrix of the system, $\rho(t)$. Recalling $\theta=2\arcsin\left(\sqrt{n_k(t)}\right)\to\sin(\theta/2)^2=n_k(t)$, we find (see App.~\ref{appendix2})

% 	\begin{equation}
% 	\operatorname{Tr}_{\text{anc}} \rho_3 = \rho(t) =
% 	%\left(
% 	%\begin{array}{cc}
% 	%\cos^2\left(\theta/2\right) & 0 \\
% 	%0 & \sin^2\left(\theta/2\right) \\
% 	%\end{array}
% 	%\right)=
% 	\left(
% 	\begin{array}{cc}
% 	1-n_k(t) & 0 \\
% 	0 & n_k(t) \\
% 	\end{array}
% 	\right)
% 	\end{equation}
% for any incident $a$ and $b$, as claimed.

\par We ran the circuit shown in Fig.~\ref{q:circuit3} on IBM's Boeblingen quantum computer\cite{ibm} for 3 different initial conditions using 3 different angles (three different $n_k$ values) for a total of 9 different circuits. We also performed quantum state tomography on the resulting density matrix, which requires measurements in the $X$, $Y$ and $Z$ bases for each circuit, for a total of 27 different runs. The circuits were optimized by hand in Qiskit\cite{Qiskit} to maximize the fidelity of the process by choosing the ideal set of qubits. The $R_y(\theta)$ gate is implemented as $U_3(\theta,-\pi/2,\pi/2)$. The connectivity of the chip allowed for an implementation without any SWAP operations. As was done above, the resulting output was corrected by Qiskit Ignis' ``pseudo-inverse'' method~\cite{Qiskit}. The data is in good agreement overall with the predicted behavior, having an average fidelity of over 99.6\% across the 9 runs with a minimum fidelity of 99.1\%.

\begin{figure*}[ht!] 
\includegraphics[width=1.5\columnwidth]{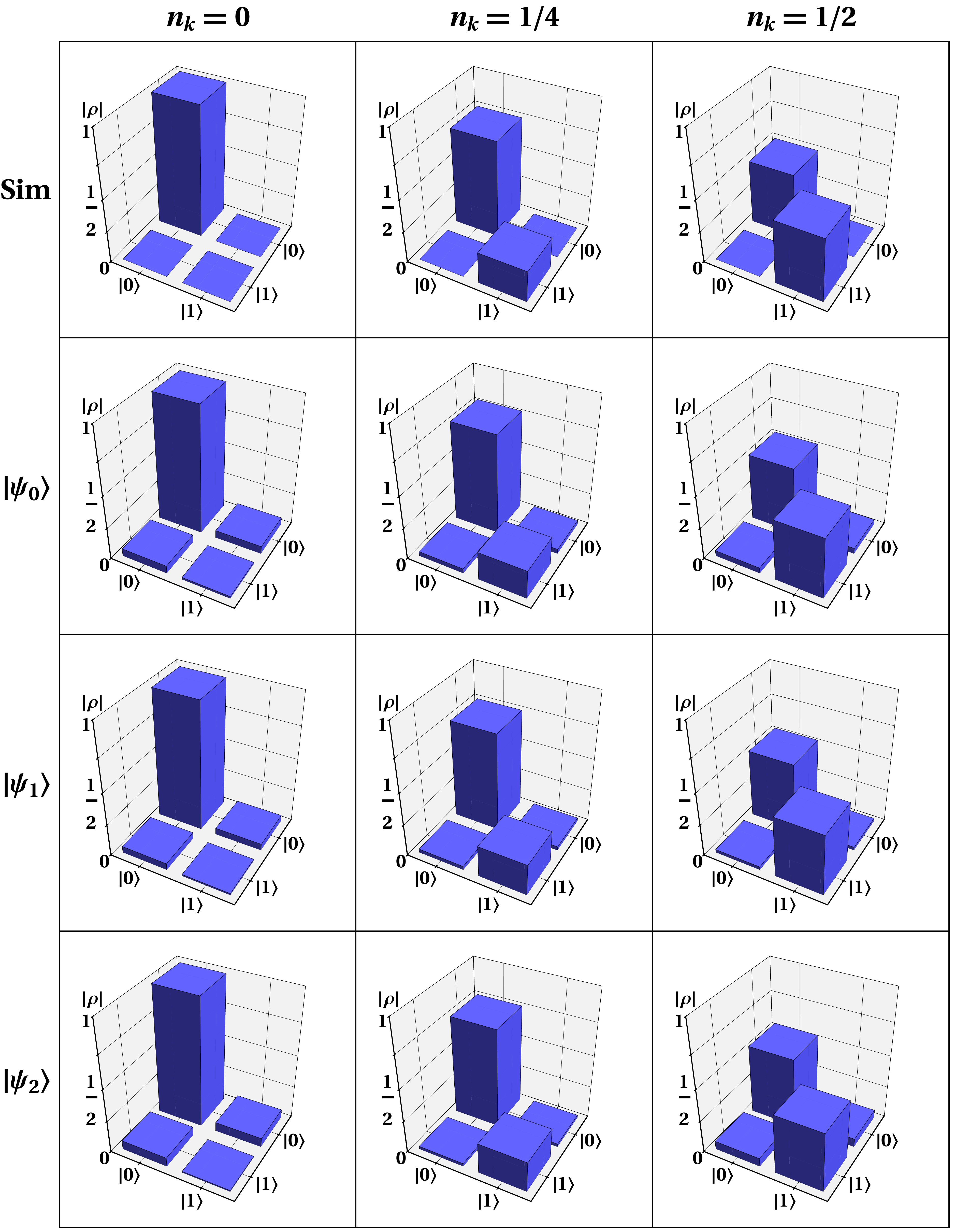}
\caption{Quantum state tomography of the simplified Kraus map circuit, as implemented on the IBM Boeblingen machine\cite{ibm}
. The amplitudes of the four density matrix elements are plotted for three different target final states, indicated by the value of $n_k$ (top label)---for a given $k$ value, the precise value for $n_k$ is determined from Eq.~\ref{mom:dist2}---here, we chose three representative values to test. The circuit is designed to work for arbitrary initial states $\rho_0$. In this work, we tested three initial states, given by: $\ket{\psi_0}=\ket{0}$, $\ket{\psi_1}=H\ket{0}=\frac{1}{\sqrt{2}}(\ket{0}+\ket{1})$, $\ket{\psi_2}=R_x(\pi/4)\ket{0}$. The measured density matrices agree well with the exact results from the simulator (extrapolated to an infinite number of shots) given in the top row, having an average fidelity of over 99.6\%. }
\label{fig:results}
\end{figure*}

The results in Fig.~\ref{fig:results} show the components of the $2\times 2$ density matrix in the standard tomography format. We plot only the amplitudes of the matrix elements, because the off-diagonal elements of the exact result vanish (hence the measured phase of those elements varies widely due to noise, and represents unimportant, but distracting, errors). 
%\LK{A typical solution to this problem is to set a color scale such as red-gray-blue, rather than red/blue. That is, you scale color by something like phase * amplitude rather than just phase. Just a suggestion if you'd like to use colors for the phases anyway.}
The label on the left hand side indicates the state that the system was initialized in ($\rho_0=\ket{\psi_0}\bra{\psi_0}$) for a particular run.

%\onecolumngrid
%\twocolumngrid

While this result is a simple implementation of a driven-dissipative system on a quantum computer, it does show that one can run such  hybrid classical-quantum simulations even on quantum computers that have no selective reset capability. As selective resets become more widely available, more complex circuits for these types of problems will become possible (and, of course, will be needed to work on more complex systems).

\section{Conclusions and Discussions}\label{sec:V}
We are interested in the question of how to most efficiently simulate driven-dissipative systems on a quantum computer. While the work presented here does not solve that problem, it provides some important results that will help us along this path. 
In particular, we have considered a system of non-interacting electrons in a lattice driven out of equilibrium by an electric field and coupled to a bath, whose dynamics is governed by the ME in Eq.(\ref{master:gen}).
\par 
We investigated how one can simulate this master equation on a quantum computer with a hybrid classical-quantum algorithm. 
We did not focus on the general result, which is known to be a hard problem, but instead looked at the simplest concrete example which can be simulated now on NISQ machines---the case of a single qubit system.
 We did this two ways. First with a Trotterized Kraus map that can examine the transient dynamics and does not require knowledge of the NESS a priori. However, for this we need to reset the ancilla register every Trotter step. Due to the lack of native reset capabilities this protocol is prohibitively expensive on current devices. To mitigate this we use large Trotter steps, which introduce an error, and still cannot evolve all the way to the steady state. Due to the prohibitive cost of simulating the dynamics with current hardware, we introduced a second circuit that reproduces the quantum operation encoded in the long-time limit of the Kraus map. This then produces the steady-state dynamics of the driven-dissipative system. The simplified circuit requires only three qubits and three controlled gates. Both of these circuits were run on IBMQ's machines\cite{ibm}. The data are generally in excellent agreement with the exact results. 
\par 
One question that we need to discuss is how scalable is such an approach? In this work, the noninteracting nature of the problem allowed for the Kraus map to be found in an integrated form. In the more realistic scenario where the system has on-site Coulomb repulsion between electrons, the situation becomes much more complicated. Finding an analytic form for the Kraus map is no longer possible and so a Trotterized form must be found. Implementing such a map requires either a fresh ancilla register for each Trotter step or the ability to reset the ancilla register without affecting the system. 
Furthermore, given the many-body nature of the interacting system, each Trotter step is likely to become cumbersome, making the simulation much more challenging.
Our Trotterized scheme (i) could be applied to a more complicated multi-qubit system where the jump operators connect different computational basis states. However, most interesting systems are not described by such simple jump operators, and constructing an approximation or extension of our method for handling a many-body interacting problem goes beyond the scope of our current work.
\par Devising approximate methods to find and implement the Kraus map related to a master equation will be important in order to simulate an interacting driven-dissipative system on currently available (or near term) quantum machines. We plan to tackle this problem in future work.

%one must use a Trotterized version of the map in Eq.~(\ref{K:map}); this should increase the number of quantum operations needed and requires partial gate resets for the ancilla qubits. \textcolor{red}{say something about number of operators?}

\section{Acknowledgments}
This work was supported by the U.S. Department of Energy, Office of Science,
Basic Energy Sciences, Division of Materials Sciences and Engineering under Grant No. DE-SC0019469.
J.~K.~F.~was also supported by the McDevitt bequest at
Georgetown. B.~R.~was also supported by the National Science Foundation under Award DMR-1747426 (QISE-NET). We acknowledge the use of IBM Q for this work. The views expressed are those of the authors and do not reflect the official policy or position of 
IBM or the IBM Q team.

\appendix
\section{Derivation of the Master Equation}\label{appendix}
For completeness, we show the derivation of the master equation as it appears in Eq.~(\ref{master:gen}). The total Hamiltonian of the system is given by $\hat{\mathcal{H}}_{tot}^{(k)} = \hat{\mathcal{H}}^{(k)}+\hat{\mathcal{H}}_b^{(k)} + \hat{\mathcal{V}}^{(k)}$, where: 

\begin{eqnarray}
\label{ham:sys1}
\hat{\mathcal H}^{(k)} &=& -2\gamma \cos\left(k+\Omega\,t\right)\dd_k d^{\phantom{\dagger}}_{k},\\
\label{ham:bath}
\hat{\mathcal H}_b^{(k)} &=& \sum_\alpha \omega_\alpha \cd_{k\alpha}\cc_{k\alpha},\\
\label{ham:hyb1}
\hat{\mathcal V}^{(k)} &=& -g\sum_{\alpha}\dd_k \cc_{k\alpha} + \mbox{h.~c.}
\end{eqnarray}

Therefore, in our case, we have a master equation for every $k$-point, but we will omit the $k$-subscript to simplify the notation.
\par For deriving the master equation, it is useful to work within the interaction picture, where the interaction in this case is given by the bilinear hybridization term $\hat{\mathcal{V}}$. A generic operator $\hat{\mathcal{O}}$ can be written in the interaction picture as $\hat{\mathcal{O}}_I(t) \equiv \hat{U}^\dag_b(t)\otimes \hat{U}^\dag(t)\,\hat{\mathcal{O}}\,\hat{U}(t)\otimes \hat{U}_b(t)$, where $\hat{U}(t)$ and $\hat{U}_b(t)$ are respectively the time-evolution operators of the isolated system and the bath and they obey the following differential equations $i\partial_t \hat{U} = \hat{\mathcal{H}}\hat{U}$, $i\partial_t \hat{U}_b = \hat{\mathcal{H}}_b\hat{U}_b$, with initial condition $\hat{U}(0) = \mathbbm{1}$, $\hat{U}_b(0) = \mathbbm{1}$ .
\par The Von-Neumann equation for the density matrix $\hat \chi_I(t)$ of the system plus bath (in the interaction picture) reads: $\partial_t\hat{\chi}_I(t) = i \left[\hat{\chi}_I(t),\hat{\mathcal{V}}_I(t)\right]$, and can be recast in an integral formulation as follows:
\begin{equation}
\hat{\chi}_I(t) = \hat{\chi}_I(0)+i\int_0^tdt^\prime \left[\hat{\chi}_I(t^\prime),\hat{\mathcal{V}}_I(t^\prime)\right].
\end{equation}
Substituting the last equation into the Von-Neumann equation in the differential form, one obtains: 
\begin{equation}\label{eq:esatta}
\partial_t\hat{\chi}_I(t) = i[\hat{\chi}_I(0),\hat{\mathcal{V}}_I(t)]-\int_0^tdt^\prime \left[ \left[\hat{\chi}_I(t^\prime),\hat{\mathcal{V}}_I(t^\prime)\right],\hat{\mathcal{V}}_I(t)\right].
\end{equation}
If the hybridization strength is small enough, we can neglect the correlations between the system and the bath. Furthermore, if the bath is a proper thermal reservoir, it is not affected very much by the dynamics of the system and its evolution as a function of time can be neglected. Under these circumstances 
we can assume the following form of the density matrix $\hat{\chi}_I(t) = \hat{\rho}_I(t) \otimes\hat{\rho}_b(0)$, also known as the Born approximation. A further approximation that we consider consists in replacing the time dependence of $\hat{\rho}_I(t^\prime)\to\hat{\rho}_I(t)$ in the integral in Eq.~(\ref{eq:esatta}) and sending the lower extremum of the integral from zero to $-\infty$, which is known as the Markov approximation. In fact, following this prescription, we  obtain a first-order differential equation for $\hat{\rho}(t)$ with constant coefficients at equilibrium. 
\par
After making all these assumptions, setting $\hat{\rho}_b(0) = e^{-\beta\hat{\mathcal{H}}_b}$, and tracing out the bath degrees of freedom we obtain the master equation: 
\begin{equation}\label{interaction}
\partial_t{\hat{\rho}}_I(t) = -\mbox{Tr}_b\int_{-\infty}^{t}dt_1 \left[\hat{\mathcal{V}}_I(t),\left[\hat{\mathcal{V}}_I(t_1),\hat{\rho}_I(t)\otimes\hat{\rho}_b(0)\right]\right],
\end{equation}
where $\hat{\rho}_I(t) = \mbox{Tr}_b \hat{\chi}_I(t)$.
The time evolved destruction operators of the bath are given by $c^{\,}_{\alpha}(t) = \hat{U}^\dag_b(t)\,c^{\,}_{\alpha}\,\hat{U}_b(t)=e^{-i\omega_\alpha t}c^{\,}_{\alpha}$.
Therefore the operator defined in Eq.~(\ref{interaction}) takes the following form in the interaction picture:
\begin{equation}\label{hyb:inter}
\hat{\mathcal{V}}_I(t) = -g\sum_{\alpha} d^{\dag}_{}(t)c^{\, }_{\alpha} e^{-i\omega_\alpha t}+ c^{\dag}_{\alpha}d^{\, }_{}(t) e^{i\omega_\alpha t},
\end{equation}
where $d_{}(t) = \hat{U}^\dag(t)\,\hat{d}_{}\,U(t)$.
\par 
If we substitute Eq.~(\ref{hyb:inter}) into Eq.~(\ref{interaction}), after some significant algebra, we obtain the master equation in the interaction picture:
\begin{widetext}
\begin{eqnarray}\label{eq:big}
\partial_t \hat{\rho}_I(t) &=& g^2\int_{-\infty}^tdt_1 \,\mathcal{C}_p(t-t_1) \left[-d^{\,}(t)\dd(t_1)\hat{\rho}_I(t)+\dd(t_1)\hat{\rho}_I(t) d^{\,}(t_1) \right] + \mbox{h.~c.} \nonumber \\
&+&g^2\int_{-\infty}^tdt_1 \,\mathcal{C}_h(t-t_1) \left[-\dd(t)d^{\,}(t_1)\hat{\rho}_I(t)+d^{\,}(t_1)\hat{\rho}_I(t) \dd(t) \right] + \mbox{h.~c.} \, ,
\end{eqnarray}
\end{widetext}
where $\mathcal{C}_p(t) = \mbox{Tr}_b\sum_\alpha \hat{\rho}_b\cd_{k\alpha}(t)\cc_{k\alpha} $, $\mathcal{C}_h(t) = \mbox{Tr}_b\sum_\alpha \hat{\rho}_b\cc_{k\alpha}(t)\cd_{k\alpha}$ are the correlation functions of the bath. We choose a half-filled bath, which implies that $\mathcal{C}_p(t) =\mathcal{C}_h(t)$ due to particle-hole symmetry.
\par
In order to solve the master equation, we have to specify the form of the correlation function of the bath.
Given the simple form of the bath Hamiltonian in Eq.~(\ref{ham:bath}), the correlation function is $k$-independent and can be calculated analytically. In particular, in the limit of an infinite bandwidth with a flat density of states [$N(\epsilon) \equiv \sum_\alpha \delta(\epsilon-\omega_\alpha)\sim N(0)$], we find that
\begin{equation}\label{corr:func_app}
\mathcal{C}_p(t) =\pi N(0) \left[\delta(t)-\frac{i}{\beta}\mbox{PV}\,\mbox{cosech}\left(\frac{\pi t}{\beta}\right)\right].
\end{equation}
Here, PV denotes the principal value.
\par 
Let us discuss some peculiarities arising from considering a flat density of states. First, in this limit the correlation function does not depend on the chemical potential of the bath, that we can set to an arbitrary number since the beginning (zero in our case). Second, it leads to divergences when one wants to calculate observables of the bath as for instance the density. It would have been more formally correct to consider an half-filled normalized density of states as a Lorentzian or a uniform box of length $W$, but this would also complicate our analytical calculations. Furthermore, this does not give appreciable deviations from what we have calculated in the limit of a flat DOS, in the case of a large bandwidth is considered, as we checked numerically. 
\par So far, we used the interaction picture. For obtaining the master equation for the density matrix in the Schr\"odinger frame, we have to ``erase" the time evolution on the system operators, that is $\hat{\rho}(t) = \hat{U}(t)\hat{\rho}_I(t) \hat{U}^\dag(t)$ and if we do so in Eq.~(\ref{eq:big}), we obtain the following equation:
\begin{eqnarray}\label{master:gen1}
\partial_t\hat{\rho}(t) &=& -i[H,\hat{\rho}(t)]
\nonumber \\
 &&-d^{\,}_{}\mathcal{D}^{(p)\dag}_{}(t)\,\hat{\rho}(t) + \mathcal{D}^{(p)\dag}_{}(t)\,\hat{\rho}(t)\,d_{} 
+ \mbox{h.~c.}
 \nonumber \\
&& 
-d^\dag_{}\mathcal{D}_{}^{(h)}(t)\,\hat{\rho}(t) + \mathcal{D}^{(h)}_{}(t)\,\hat{\rho}(t)\,d^{\dag}_{}+ \mbox{h.~c.}, \nonumber \\
\end{eqnarray}
where we defined the following operators:
\begin{eqnarray} \label{d:dag:p}
\mathcal{D}_{}^{(p)\dag}(t) &\equiv& g^2\int_{-\infty}^{t}dt_1 \,\mathcal{C}_p(t-t_1)\mathcal{D}_{}^\dag(t,t_1), \\ 
\label{d:h}
\mathcal{D}_{}^{(h)\,}(t) &\equiv& g^2\int_{-\infty}^{t}dt_1 \,\mathcal{C}_h(t-t_1)\mathcal{D}_{}(t,t_1).
\end{eqnarray}
Here $\mathcal{D}_{}^\dag(t,t_1) \equiv \hat{U}(t)\hat{U}^\dag(t_1)\,d\,\hat{U}(t_1)\hat{U}^\dag(t)$.
It is worthwhile to note that the two operators defined in Eqs.~(\ref{d:dag:p}) and (\ref{d:h}) are not conjugates of each other (because of the finite imaginary part in $\mathcal{C}_{p/h}$). 
\par 
Therefore, since we do not neglect the structure of the correlation functions of the bath, we have in principle to construct the new operators defined in Eqs.~(\ref{d:dag:p}) and (\ref{d:h}).
If the Hamiltonian of the system does not depend explicitly on time, the operators in Eqs.~(\ref{d:dag:p}) and (\ref{d:h}) would not depend on time and a similar expression can be found as those found in the studies of transport in quantum dots\cite{Harbola2006,Welack2006,Welack2008,Welack2009}.
In our case, the time dependence of the Hamiltonian and a nontrivial structure of the correlation function [see Eq.~(\ref{corr:func})] yields time-dependent operators. 
\par 
A further simplification comes from the one-body nature of our problem. From now on, we will introduce again the $k$ subscript in our notation. In this case, the time-dependent annihilation operator can be written as $d_k(t) = e^{-i F_k(t)}d $, where $F_k(t) = -2\gamma\left[\sin(k+\Omega t) - \sin(k)\right]/\Omega$. Therefore 
\begin{equation}
\mathcal{D}_k(t,t_1) = e^{-i[ f_k(t_1)-f_k(t)]}\,d,
\end{equation}
with $f_k(t) = \sin(k + \Omega t)/\Omega$, 
and the RME reads:
\begin{eqnarray}
\partial_t\hat{\rho}_k&=&\mbox{Re}\,a^{\phantom{\dagger}}_k(t)\left[2\dd_k\hat{\rho}^{\phantom{\dagger}}_k d^{\phantom{\dagger}}_k-\left\{d^{\phantom{\dagger}}_k\dd_k,\hat{\rho}^{\phantom{\dagger}}_k \right\}\right] \nonumber \\
&+&\mbox{Re}A^{\phantom{\dagger}}_k(t)\left[2d_k\hat{\rho}^{\phantom{\dagger}}_k \dd_k-\left\{\dd_k d^{\phantom{\dagger}}_k,\hat{\rho}^{\phantom{\dagger}}_k \right\}\right],
\end{eqnarray}
where we introduced the time and momentum dependent coefficients: 
\begin{eqnarray}
\label{coeffs:app}
a_k(t) &=& g^2 \exp\left[-if_k(t) \right]\int_{-\infty}^0dt_1 \mathcal{C}_p(-t_1) \exp\left[if_k(t+t_1) \right], \nonumber \\
A_k(t) &=&  g^2 \exp\left[if_k(t) \right]\int_{-\infty}^0dt_1 \mathcal{C}_h(-t_1) \exp\left[-if_k(t+t_1) \right], \nonumber \\
\label{eq:coeffs}
\end{eqnarray}
and we used the fact that $\mbox{Im}(a_k(t) - A_k(t)) = 0$.

\subsection{Analytic expressions of \texorpdfstring{$\boldsymbol{n(k_m)}$}{n(km)} and
\texorpdfstring{$\boldsymbol{\langle J\rangle}$}{<J>} in the RME and LME} 
\par Our next step is to simplify the coefficients $a_k(t)$ and $A_k(t)$. Employing the standard Bessel function identity
\begin{equation}
 \exp[i f_k(t)] = \sum_{\ell=-\infty}^{+\infty}J_\ell\left (\frac{2\gamma}{\Omega}\right )\exp[-i\ell (k+\Omega t)], 
\end{equation}
allows us to re-express the coefficients in Eq.~(\ref{coeffs}) as the following:
\begin{eqnarray}
a_k(t) &=& \Gamma\sum_{\ell\ell^\prime} J_\ell\left(\frac{2\gamma}{\Omega}\right)J_{\ell^\prime}\left(\frac{2\gamma}{\Omega}\right)\mathcal{F}(\Omega \ell)\, e^{-i(\ell-\ell^\prime)(k + \Omega\,t)}, \nonumber \\
A_k(t) &=& \Gamma\sum_{\ell\ell^\prime} J_\ell\left(\frac{2\gamma}{\Omega}\right)J_{\ell^\prime}\left(\frac{2\gamma}{\Omega}\right)\mathcal{F}(-\Omega \ell)\, e^{i(\ell-\ell^\prime)(k + \Omega\,t)}. \nonumber \\
\end{eqnarray}
We choose $\Gamma = \pi g^2 N(0)$ and 
\begin{equation}\label{Fourier:Transf}
\mathcal{F}(x) = n_F(-x)+\frac{i}{\pi}\mbox{Re}\,\psi\left(\frac{1}{2}-i\frac{\beta\, x}{2\pi}\right),
\end{equation}
with $\psi(z)$ being the digamma function.

 From Eq.~(\ref{master:gen}), we can obtain a differential equation for the momentum distribution function $n_k(t) = \mbox{Tr}\left(\hat{\rho}^{\phantom{\dagger}}_k(t) \dd_k\,d^{\,}_k\right)$, that reads: 
\begin{equation}\label{mom:dist1} 
\dot{n}_k = -2\Gamma n_k + 2\,\mbox{Re}\, a_k(t).
\end{equation}
Since this is a first-order linear differential equation, it can be immediately integrated to yield its solution, which given by
\begin{equation}\label{mom:dist2}
n_k(t) = e^{-2\Gamma(t-t_0)}n_k(t_0) + \int_{t_0}^{t}ds\,e^{-2\Gamma(t-s)}\,2\,\mbox{Re}\,{a_k(s)}.
\end{equation}
In the limit of $t_0\to-\infty$, Eq.~(\ref{mom:dist2}) becomes
\begin{equation}\label{mom:dist}
n_k(t) = 2\,\Gamma\,\mbox{Re}\,\sum_{\ell\ell^\prime} \frac{J_\ell\left(\frac{2\gamma}{\Omega}\right)J_{\ell^\prime}\left(\frac{2\gamma}{\Omega}\right)\mathcal{F}(\Omega \ell)}{2\Gamma-i(\ell-\ell^\prime)\Omega}e^{-i(\ell-\ell^\prime)(k + \Omega\, t)}.
\end{equation}

The equation for the current is given by $\langle J\rangle  = (2\pi)^{-1}\int dk\,2 \gamma \sin(k+ \Omega \,t)n_k(t)$ and in the limit of $t\to \infty$ we have: 
\begin{eqnarray}\label{current:RME}
\langle J\rangle &=& 4\gamma\,\Gamma\,\mbox{Re}\sum_\ell 
\frac{J_\ell\left(\frac{2\gamma}{\Omega}\right)
J_{\ell+1}\left(\frac{2\gamma}{\Omega}\right)
\mathcal{F}(\Omega \ell)}{\Omega-2i\Gamma} \nonumber \\
&+&4\gamma\,\Gamma\,\mbox{Re}\sum_\ell 
\frac{J_\ell\left(\frac{2\gamma}{\Omega}\right)
J_{\ell-1}\left(\frac{2\gamma}{\Omega}\right)
\mathcal{F}(\Omega \ell)}{\Omega+2i\Gamma} \,.
\end{eqnarray}

In Figs.~(\ref{fig:mom:t},\ref{fig:mom:dist}), where we evaluated the momentum distribution function as well as in Fig.~(\ref{fig:curr}) where we evaluated the current in the RME, we had to
set a cut-off to the maximum index $\ell$ appearing in Eqs.~(\ref{mom:dist}) and (\ref{current:RME}). This value ranges from 20 in the case of strong-driving fields to 160 in the case of our smallest finite field that is $\Omega/\gamma = 0.05$.
Now, we will derive the analytic expression for the momentum-distribution function obtained through the LME. In this case, we do not need to employ the Bessel function expansion and the integral in Eq.~(\ref{mom:dist2}) can be calculated directly. In fact, the coefficient $a_k(t) \sim \Gamma n_F\left[\epsilon_k(t)\right]$ and if we define $k_m = k + \Omega\,t$, we can rewrite Eq.~(\ref{mom:dist2}) as:
\begin{eqnarray}\label{last:eq}
    n(k_m) &\sim& \frac{2\Gamma}{\Omega}\int_{-km}^{0}dx\,e^{-\frac{2\Gamma}{\Omega}(x + k_m)}n_F[-2\gamma\cos(x)] \nonumber \\
    &+&\frac{2\Gamma}{\Omega}\sum_{n=0}^{\infty}\int_{2 \pi n}^{2 \pi(n+1)}dx\,e^{-\frac{2\Gamma}{\Omega}(x + k_m)}n_F[-2\gamma\cos(x)]. \nonumber \\
    \end{eqnarray}
In the limit of $T\to 0$, the integrals in Eq.~(\ref{last:eq}) can be calculated straightforwardly and for $k_m\in [0,2\pi]$ we obtain:
\begin{equation}
    n(k_m) \sim  e^{-\frac{2\Gamma}{\Omega}k_m}\left[\mathcal{I}(k_m) -\frac{1}{2} \text{sech}\left(\pi  \frac{\Gamma}{\Omega} \right)\right],   
\end{equation}
where:
\begin{equation}
\mathcal{I}(x) = 
\begin{cases}
 e^{2  \frac{\Gamma}{\Omega}  x} & x<\frac{\pi }{2} \\
 e^{\pi   \frac{\Gamma}{\Omega} } & \frac{\pi }{2}\leq x\leq \frac{3 \pi }{2} \\
 e^{\pi   \frac{\Gamma}{\Omega} }-e^{3 \pi   \frac{\Gamma}{\Omega} }+e^{2  \frac{\Gamma}{\Omega}  x} & \frac{3 \pi }{2}<x\leq 2 \pi .
\end{cases}
\end{equation}
We note that the solutions of the LME depend solely on the ratio $\Gamma/\Omega$, while the solutions of the RME do not have this property. This is due to the fact that the coefficients $a_k(t)$ depend on $\Omega$ in a non-trivial way, while in the case of the LME the $a_k(t)$ depends on $\Omega$ only through $k_m$. 
\par
The expression of the current significantly simplifies as well if one calculates it from the LME. In fact, we can rewrite the  equation of motion of the momentum distribution function as
\begin{equation}\label{LME:k_m}
\frac{d n}{d k_m}(k_m) = \frac{2 \Gamma}{\Omega}\left(-n(k_m) + n_F(k_m) \right).
\end{equation}

In order to obtain the current, we first multiply both sides of Eq.~(\ref{LME:k_m}) times $2\gamma\,\sin(k_m)$ and integrate over $k$ and then we multiply the same equation times $2 \gamma \cos(k_m)$ and integrate again. In this way, we obtain the following set of equations:
\begin{eqnarray}
\big<\tilde{J}\big> &=&  \frac{2\Gamma}{\Omega}\big<J\big> \nonumber \\
\big<J\big> &=& -\frac{2\Gamma}{\Omega}\big<\tilde{J}\big> +\frac{4\gamma \Gamma }{\Omega}\,I,
\end{eqnarray}
where, $\big<\tilde{J}\big>  = (2\pi)^{-1}\int dk_m\,2\gamma\cos(k_m)\,n({k_m})$ , $I = (2\pi)^{-1}\int dk\,n_F(\epsilon_k)\,2\gamma \cos(k)$. At zero temperature $I = 2\gamma/\pi$ and the current obtained using the LME reads:
\begin{equation}\label{current:LME}
\left<J\right>\sim \frac{2 \gamma}{\pi}\frac{2\Gamma/\Omega}{1 + \left(2\Gamma/\Omega\right)^2}.
\end{equation}
The same formula has been found in Ref.~\onlinecite{Han2012} using the Keldysh formalism in the limit $\Omega \ll 1$ and $\Gamma \ll 1$, which is consistent with the Born approximation and the expansion in Eq.~(\ref{eq:expansion}) that we performed to obtain the LME.

\section{Derivation of the linear map \texorpdfstring{$\boldsymbol\phi_t$}{\textphi}}\label{linear:map}
Here we explicitly solve for the linear map $\phi_t$ in its matrix representation $F_{rs} = \mbox{Tr}\left({\sigma_r \phi_t(\sigma_s)}\right)$, which is defined in Sec.~\ref{sec:int:map}, where 
$\sigma_{0} = \mathbbm{1}_{2\times 2}/\sqrt{2}$,
 $\sigma_{1} = \sigma^z/\sqrt{2}$,
 $\sigma_{2} = \sigma^x/\sqrt{2}$,
 $\sigma_{3} = \sigma^y/\sqrt{2}$.
 For this purpose, let us construct the matrix representation of the map $\Lambda_t(\rho) = \dot{\rho}$, which is defined as $L_{rs} = \mbox{Tr}{\sigma_r\Lambda_t(\sigma_s)}$.  As a matter of fact, $L$ has a block diagonal form where the blocks are defined in the parallel subspace, spanned by $\mathbbm{1}_{2\times 2}$ and $\sigma^z$ and the transverse subspace, spanned by $\sigma^x$ and $\sigma^y$. The matrix representation of $\Lambda_t$ in the two different channels reads:
 \label{para}
\begin{equation}
L_{\parallel} = 
-2\left(\begin{array}{cc}
0&0 \\
\mbox{Re}[a_k(t)-A_k(t)] & \Gamma
\end{array}
\right),
\end{equation}
\label{perp}
\begin{equation}
L_{\perp} = 
- \Gamma\left(\begin{array}{cc}
1&0 \\
0 & 1
\end{array}
\right),
\end{equation}
where we used the fact that $\mbox{Re}[a_k(t)+A_k(t)] = \Gamma$. We note that these two properties are satisfied in the  infinite flat bandwidth limit, and they would not hold for a non-trivial choice of the bath electrons DOS.
Therefore, the dynamics of the two channels is totally decoupled. We further note, that $F_{\perp}$  vanishes when $t\to\infty$. Hence, the information about the steady state is fully contained in the parallel channel, whose time evolution is given by $\partial_t F_\parallel = L_\parallel F_\parallel $, with the initial condition $F_\parallel(0)= \mathbbm{1}_{2\times 2}$ and its solution reads:
\begin{equation}
F_{\parallel} = 
\left(\begin{array}{cc}
1&0\\
\Phi(t)  & e^{-2\Gamma t}
\end{array}
\right),
\end{equation}
where 
\begin{eqnarray}
\Phi(t) = -2\mbox{Re}\int_{0}^tdt_1 [a_k(t_1)-A_k(t_1)]e^{2 \Gamma (t_1-t)}.
\end{eqnarray}
Using the fact that $\mbox{Re}\left[a_k(t)-A_k(t)\right] = \mbox{Re}\left[a_k(t)-A_k^*(t)\right] = 2\mbox{Re}[a_k(t)]- 2\Gamma$, substituting this into the last equation, for long enough time we have
\begin{equation}
\Phi(t) = -4\mbox{Re}\int_{-\infty}^tdt_1 a_k(t_1)e^{2 \Gamma (t_1-t)}+\,1 = -2n_k(t) +1,
\end{equation}
where  we substituted $t_0 = 0\to-\infty$ and used Eq.~(\ref{mom:dist2}).
\par With this information, we are now able  to calculate the Choi matrix $S_{ab} = \sum_{r=0}^3\sum_{s=0}^3F_{sr}\mbox{Tr}\left[\sigma_r\sigma_a\sigma_s\sigma_b\right]$, whose orthogonal decomposition yields the Kraus operators defined in Eq.~(\ref{kraus_map}).

\bibliography{biblio} 

\end{document}